\documentclass[10pt,journal]{IEEEtran}
\usepackage{color}
\usepackage{cite}
\usepackage{subfigure}

\ifCLASSINFOpdf
\usepackage[pdftex]{graphicx}
\graphicspath{{../pdf/}{../jpeg/}}
\DeclareGraphicsExtensions{.pdf,.jpeg,.png}
\else
\usepackage[dvips]{graphicx}
\graphicspath{{../eps/}}
\DeclareGraphicsExtensions{.eps}
\fi
\usepackage{makecell,multirow,diagbox}
\usepackage{pifont}
\usepackage{enumerate}
\usepackage{amsfonts}
\usepackage{booktabs}
\usepackage{algorithmic}
\usepackage{algorithm}
\usepackage[cmex10]{amsmath}

\usepackage{mathabx}
\usepackage{mathrsfs}
\usepackage{subeqnarray}
\usepackage{cases}
\usepackage{color}
\begin{document}
%
\title{Secure Collaborative Computation Offloading and Resource Allocation in Cache-Assisted Ultra-Dense IoT Networks With Multi-Slope Channels}
\author{Tianqing~Zhou,~
        Bobo~Wang,~
        Dong~Qin,~
        Xuefang~Nie,~
        Nan~Jiang,~
        and Chunguo~Li~
\thanks{This work was supported by National Natural Science Foundation of China under Grant Nos. 62261020, 62171119, 62361026 and 62461036, Jiangxi Provincial Natural Science Foundation under Grant Nos. 20232ACB212005, 20224BAB202001 and 20232BAB202019, Key research and development plan of Jiangsu Province under Grant No. BE2021013-3.} %
\thanks{T. Zhou, X. Nie and N. Jiang are with the School of Information and Software Engineering, East China Jiaotong University, Nanchang 330013, China (email: zhoutian930@163.com; Xuefangnie@163.com; jiangnan1018@gmail.com).}
\thanks{B. Wang is with the College of Electronic and Information Engineering, Huaibei Institute of Technology, Huaibei, 235000, China (email: wangbobo$\_$ecjtu@163.com).}
\thanks{D. Qin is with School of Information Engineering, Nanchang University, Nanchang 330031, China (e-mail: qindong@ncu.edu.cn).}
\thanks{C. Li is with School of Information Science and Engineering, Southeast University, Nanjing 210096, China (email: chunguoli@seu.edu.cn).}}
\maketitle

\begin{abstract}
Cache-assisted ultra-dense mobile edge computing (MEC) networks are a promising solution for meeting the increasing demands of numerous Internet-of-Things mobile devices (IMDs). To address the complex interferences caused by small base stations (SBSs) deployed densely in such networks, this paper explores the combination of orthogonal frequency division multiple access (OFDMA), non-orthogonal multiple access (NOMA), and base station (BS) clustering. Additionally, security measures are introduced to protect IMDs' tasks offloaded to BSs from potential eavesdropping and malicious attacks. As for such a network framework, a computation offloading scheme is proposed to minimize IMDs' energy consumption while considering constraints such as delay, power, computing resources, and security costs, optimizing channel selections, task execution decisions, device associations, power controls, security service assignments, and computing resource allocations. To solve the formulated problem efficiently, we develop a further improved hierarchical adaptive search (FIHAS) algorithm, giving some insights into its parallel implementation, computation complexity, and convergence. Simulation results demonstrate that the proposed algorithms can achieve lower total energy consumption and delay compared to other algorithms when strict latency and cost constraints are imposed.
\end{abstract}


\IEEEpeerreviewmaketitle

\section{Introduction}

\IEEEPARstart{W}{ith} the rapid development of mobile Internet and Internet-of-Things (IoT) technologies, a growing number of various computing-intensive and delay-sensitive applications have emerged \cite{J.Zhao2024,Q.Zhang2024}. However, IoT mobile devices (IMDs) with limited battery capacity and computational resources are not efficient in fulfilling the requirements of these applications. To reduce the workloads and energy consumption of IMDs (users), mobile edge computing (MEC) supports IMDs in completely or partially offloading their tasks to some nearby edge servers for computing \cite{FGuo2018Dec,J.Zhao2024Mar}.
\par
To greatly reduce offloading time and energy consumption, and reduce resource consumption, mobile edge caching is commonly seen as a promising strategy and has been deployed on edge servers, which supports IMDs in caching popular applications/services/tasks in advance \cite{TZhou2023Mar}. To reduce the distance between IMDs and edge servers, thereby enhancing channel gains and improving frequency-spectrum efficiency, the widespread adoption of ultra-dense small base stations (SBSs) is being promoted. In this setup, edge servers are integrated into all BSs.
\par
Although cache-assisted ultra-dense MEC networks may greatly enhance the service coverage and reduce the energy consumed by IMDs, there exist some other problems arise from computation offloading. Firstly, as the computation tasks of IMDs are transmitted to the edge servers for computing, the transmission energy consumption and latency, and edge computing energy consumption and latency yield inevitably. Secondly, edge servers, positioned at the perimeter of networks and in proximity to potential attackers, are susceptible to malicious assaults. As a result, mobile devices (MDs) will unavoidably encounter supplementary computational burdens of implementing certain security measures, leading to increased computational latency and energy expenditure. Clearly, within the constraints of limited network resources, a prevailing issue is how to safeguard user data while minimizing energy consumption in cache-assisted ultra-dense MEC networks.

\subsection{Related Work}
In recent years, substantial efforts have been dedicated to designing computation offloading mechanisms for ultra-dense networks. These mechanisms can be broadly categorized into two groups based on their respective offloading modes.
\par
In the first group, the non-cooperative mode was considered, which should be the most frequently used in existing work. In \cite{FGuo2018Dec}, a hierarchical computation offloading algorithm, integrating genetic algorithm (GA) and particle swarm optimization (PSO), was devised to minimize the overall energy consumption of users in dense small-cell networks. This approach jointly optimized spectrum usage, power allocations, computation offloading decisions, and resource distributions. In \cite{TZou2021Oct}, an enhanced computation offloading algorithm, leveraging GA and PSO, was designed to reduce network-wide energy consumption in ultra-dense heterogeneous networks while adhering to users' latency constraints. It jointly optimized computation offloading decisions, power adjustment, and resource allocations. In \cite{HGuo2018Aug}, a heuristic greedy offloading strategy was introduced to address the challenge of minimizing total energy consumption in ultra-dense multi-task networks under the processing time limitations of different tasks. In \cite{ChengQQ2020}, a mean-field deep deterministic policy gradient (DDPG) algorithm was devised to minimize local energy consumption and task delay in NOMA(non-orthogonal multiple access)-enabled ultra-dense networks, operating within rate and power constraints. \cite{RZheng2021Oct} proposed a mean-field game-based offloading algorithm aimed at reducing computation and transmission expenses in ultra-dense cloud-radio access networks. \cite{LLi2021Mar} presented an iterative optimization algorithm that combined user clustering matching with a mean-field DDPG approach, minimizing users' computation costs in NOMA-enabled ultra-dense networks while guaranteeing constraints of task delay and transmission rate. In \cite{FLi2022May}, a second-price auction scheme was designed to tackle the problems of spectrum sharing and edge computation offloading for ultra-dense networks based on software-defined networking. In \cite{RZhang2022Apr}, a calibrated contextual bandit learning algorithm was developed to minimize the long-term average task delay of all users for ultra-dense networks, which optimized the computation offloading decision. In \cite{YLu2022Sep}, the interior point method and GA were employed to address a system cost minimization problem in ultra-dense networks, constrained by limited computation resources, channel resources, and battery capacity. Such an approach jointly optimized channel selections, task offloading, and resource scheduling.
\par
In the second group, the cooperative mode was considered to fully utilize network computing resources. In \cite{YDai2018Dec}, a multi-task, multi-step offloading algorithm was designed to reduce network-wide energy consumption in ultra-dense networks, jointly optimizing user association decisions, offloading decisions, local transmission power, and computation resources while adhering to users' latency constraints. Additionally, \cite{TZhou2022Oct} introduced an advanced hierarchical adaptive search algorithm to address the minimization of network-wide energy consumption in ultra-dense multi-task IoT networks. This algorithm worked within the framework of proportional computation resource allocations and users' latency constraints, achieving joint optimization of computation offloading, resource allocations, and device associations.
\par
It is easy to find that most of the above-mentioned efforts may just concentrate on the orthogonal frequency division multiple access (OFDMA), which will result in extremely low resource utilization efficiency. In addition, most of them investigated single-task systems, and few of them introduced edge caching and security measures.
\par
To substantially reduce offloading duration and energy usage, edge caching has been incorporated into ultra-dense MEC networks. A two-timescale deep reinforcement learning strategy was developed in \cite{SYu2021Feb} to minimize the long-term execution time of edge devices and the storage consumption of edge servers within a suggested intelligent ultra-dense edge computing system. This strategy concurrently optimized service caching arrangements, computation offloading decisions, and resource allocations. Additionally, \cite{ZChen2023} introduced a method that jointly optimized computation offloading decisions, cooperative service caching, and resource allocations, aiming to reduce the total weighted energy consumption of off-grid SBSs and all mobile users in ultra-dense energy-harvesting networks. It is noteworthy that the aforementioned two works considered different time slots for edge computing and caching. They may be not highly efficient since some popular applications/services/tasks can often be cached in advance according to their popularity in reality. Moreover, they didn't protect offloaded data using necessary security measures.
\par
To protect user data during computation offloading, security measures have attracted increasing attention in (ultra-dense) MEC networks. Considering that it may be very difficult to prevent user data from being attacked/leaked by cooperative attackers/eavesdroppers in the physical layer security techniques \cite{SLiu2022Jun, MWu2024Feb}, security service assignments have been widely investigated in existing efforts. In \cite{IElgendy2019}, a heuristic algorithm was formulated to reduce the system-wide processing time and energy consumption by jointly optimizing security decision policies, computation offloading decisions, and resource allocations. Meanwhile, \cite{MZahed2020} focused on jointly optimizing security service assignments, cooperative task offloading, and caching to minimize the total cost quantified as a blend of security breach costs and energy consumption in an IoT system. Furthermore, \cite{IAElgendy2020Dec} introduced an efficient and secure multi-task computation offloading algorithm aimed at minimizing the weighted total energy consumption while satisfying users' latency constraints, not only considering offloading policies but also incorporating data compression, task security decisions, and resource allocations. In \cite{IAElgendy2021Jan}, offloading decisions were dynamically made based on factors such as energy consumption, execution time, CPU usage, and memory utilization. Additionally, a new security layer was added to safeguard data transmitted to the cloud. Drawing inspiration from the secure offloading mechanisms in MEC networks mentioned earlier, Zhou \textit{et al.} in \cite{TZhou2024Feb} worked on channel selections, multistep computation offloading, device associations, power controls, and security service assignments for ultra-dense multi-task MEC networks, designed enhanced whale optimization algorithm (IWOA), aiming to minimize the energy consumption of all mobile devices (MDs).
\par
Inspired by the efforts made in \cite{MZahed2020}, our focus in this paper is on simultaneously optimizing device associations, task execution decisions, channel selections, power controls, security service assignments, and computing resource allocations, with the ultimate goal of reducing the energy consumption of MDs. However, such a consideration is significantly different from the work in \cite{MZahed2020}. Unlike the spectrum utilization manner (i.e., OFDMA) in \cite{MZahed2020}, we utilize both OFDMA and NOMA to enhance the resource utilization, which is similar to the consideration in \cite {TZhou2024Feb}. However, unlike the partial multistep offloading in \cite {TZhou2024Feb}, we consider binary collaborative offloading for a cache-assisted system. In addition, unlike the network models used in the aforementioned literature, we consider cache-assisted ultra-dense MEC networks with a more realistic and reliable channel model (i.e., a general multi-slope channel model) \cite{WTeng2021Spet}.

\subsection{Contributions and Organization}
In this paper, addressing the demand for green and secure communications for IMDs in cache-assisted ultra-dense MEC networks, and employing a more realistic and dependable channel model, we undertake a joint optimization of device associations, task execution decisions, channel selections, power controls, security service assignments, and computing resource allocations. The primary work and contributions of this paper can be concisely summarized as follows.
\begin{enumerate}[1)]
\item \textit{Secure collaborative computation offloading in cache-assisted ultra-dense MEC networks with multi-slope channels:} To the best of our knowledge, there is scarcely any research into the computation offloading in ultra-dense MEC networks with a general multi-slope channel model, and no studies have explored computation offloading for such networks with cache assistance. Additionally, the investigation of secure collaborative computation offloading in cache-assisted ultra-dense MEC networks featuring multi-slope channels, and pre-caching applications/services/tasks based on popularity, appears to be a completely novel area of research.
\item \textit{Joint optimization of secure collaborative computation offloading and resources in cache-assisted ultra-dense MEC networks with multi-slope channels:} To satisfy the requirements for green and secure communications for IMDs in cache-assisted ultra-dense MEC networks with multi-slope channels, we consider a joint optimization of device associations, channel selections, task execution decisions, computing resource allocations, security service assignments, and power controls. This optimization is conducted under constraints related to delay, power, computing resources, and security costs, with the aim of minimizing the energy consumption of IMDs. To our knowledge, this type of joint optimization constitutes a novel investigation in such networks.
\item \textit{New algorithm designed for joint optimization of secure collaborative computation offloading and resources in cache-assisted ultra-dense MEC networks with multi-slope channels:} Upon direct observation, it is evident that the ultimately formulated problem takes on a nonlinear and mixed-integer form. To solve this problem, we develop a high-performance algorithm by further improving existing IHAS (improved hierarchical adaptive search) \cite{TZhou2022Oct}, which is regarded as further improved HAS (FIHAS). Specifically, we try to improve the adaptive diversity-guided GA (ADGGA) in IHAS, establishing improved ADGGA (IADGGA). In IADGGA, we consider new highly effective crossover and mutation probabilities to search the feasible solution space of the formulated problem more fully, which can further enhance the population diversity and avoid premature convergence. In addition, unlike ADGGA, at the latter stage of iteration of IADGGA, the individuals with high similarity and low fitness are eliminated and then regenerated using the rules of population initialization. Such an operation can further enhance the population diversity to achieve better solutions.
\item \textit{The analyses of convergence, computation complexity, parallel implementation, and simulation:} With respect to the developed algorithms, we initially offer thorough analyses focusing on convergence, computational complexity, and parallel implementation. Within these analyses, particularly those concerning computational complexity and parallel implementation, we introduce highly efficient and skillful operations aimed at reducing computational complexity and simplifying practical implementation. Subsequently, we assess the effectiveness of our developed algorithms by introducing additional algorithms for comparison in the simulation.
\end{enumerate}
\par
The remainder of this paper is structured as follows. Section \ref{sec2} presents an overview of the system model, encompassing the network model, communication model, caching and computing models, as well as the security model. In Section \ref{sec3}, we formulate a problem aimed at minimizing the energy consumption of IMDs, which is subsequently addressed by the FIHAS algorithm in Section \ref{sec4}. Section \ref{sec5} delves into the convergence and complexity analyses of the FIHAS algorithm. Section \ref{sec6} shows the simulation results along with corresponding analyses. Lastly, Section \ref{sec7} offers some conclusions and suggestions for future work.
\begin{figure}[!t]
\centering
\centerline{\includegraphics[width=3.6in]{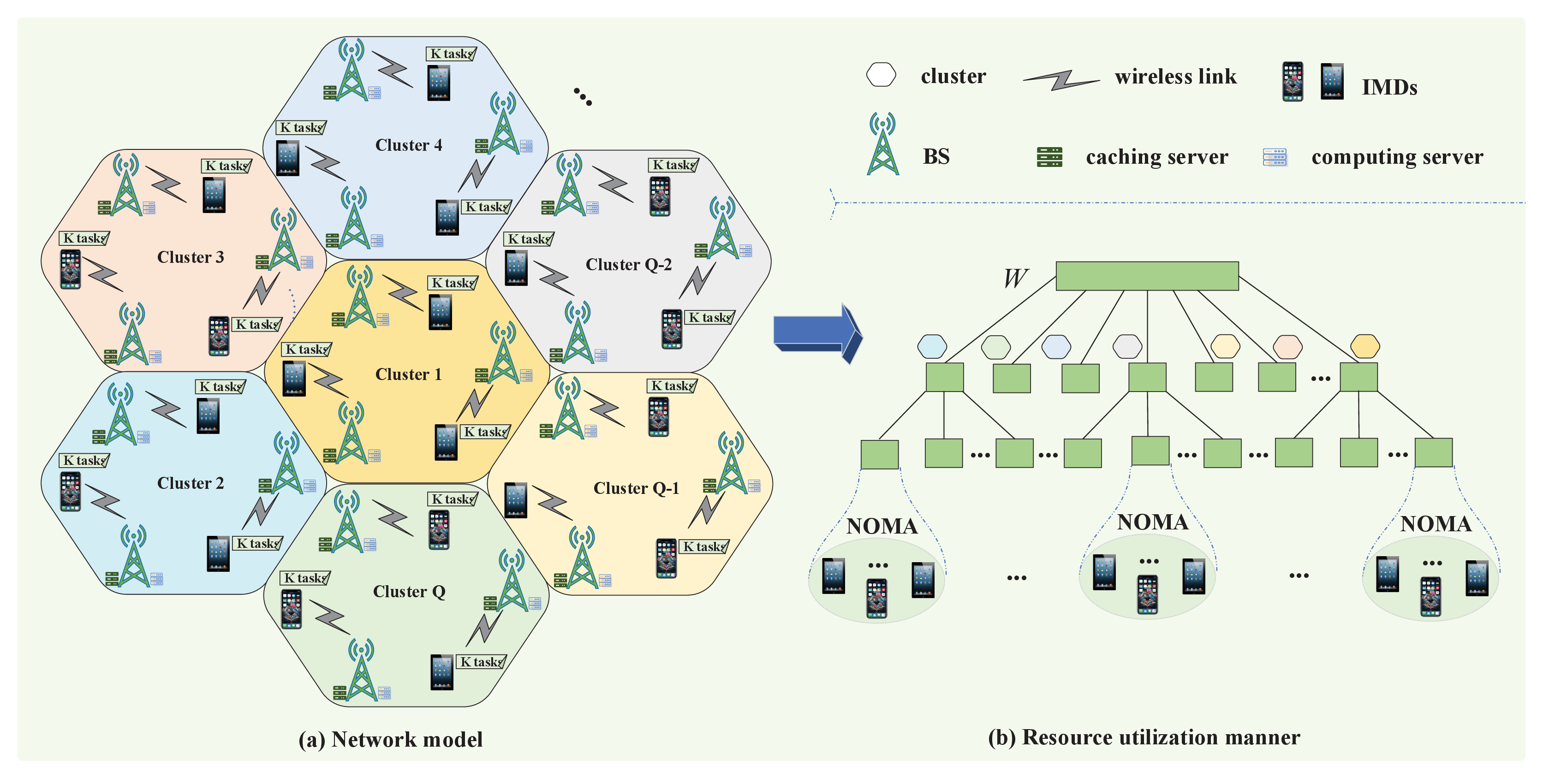}}
\caption{Cache-assisted ultra-dense MEC networks.}
\label{fig1}
\end{figure}
\section{System Model}\label{sec2}
 \subsection{Network Model}
In this paper, we concentrate on cache-assisted ultra-dense MEC networks featuring multi-slope channels, as illustrated in Fig.\ref{fig1}. In such networks, there are $N$ SBSs (BSs), indexed from 1 to $N$ within the set $\mathcal{N}$. Each SBS is equipped with both caching and computing servers. Additionally, there exist $K$ IMDs, indexed from 1 to $K$ within the set $\mathcal{K}$. Each IMD possesses $M$ types of computation tasks, indexed from 1 to $M$ within the set $\mathcal{M}$.
\par
Each type of tasks can be processed by an IMD or offloaded to edge servers for computing. When the latter manner is adopted, these tasks require confidential treatment, and the security service assignments need to be done for them. To mitigate network interference and enhance resource utilization, BSs are divided into $Q$ clusters using K-means approach according to their physical locations. We assume that different clusters utilize distinct frequency spectrums, but BSs at the same cluster have $S$ identical subchannels (frequency bands) indexed from 1 to $S$ within the set $\mathcal{S}$, where $S=R({W}/{\left( wQ \right)})$; $R(b)$ is a rounding function with respect to $b$; $W$ is system bandwidth, and $w$ is the one of a subchannel. These subchannels can be used by IMDs associated with BSs at the same cluster through NOMA manner. In such a framework, we additionally assume that each IMD can only be connected (associated) to one BS, it can utilize only one subchannel of associated BS through NOMA manner. It is evident that inter-cluster interferences don't exist, but intra-cluster interferences exist.

\subsection{Communication Model}
Due to the involvement of IMD (user) associations in computation offloading, which often occurs over large-scale time slots \cite{WTeng2021Spet}, it is essential to utilize slow fading channels instead of fast ones. This implies that the impact of fast fading needs to be averaged out in the channel model. Furthermore, as demonstrated in \cite{WTeng2021Spet}, block fading should also be averaged across different fading blocks within a single user association period. To accurately represent the pathloss in ultra-dense networks, it is recommended to adopt a comprehensive multi-slope channel model that incorporates both LoS (line-of-sight) and NLoS (non-line-of-sight) paths. Specifically, the channel gain ${{h}_{n,k}}$ between BS $n$ and IMD $k$ comprises $J$ (path) components denoted as $\{ \dddot{h}_{n,k,1},\dddot{h}_{n,k,2},\cdots ,\dddot{h}_{n,k,J} \}$. Mathematically, it is
\begin{equation}\label{eq1}
{{h}_{n,k}}=\left\{ \begin{split}
  & \dddot{h}_{n,k,1},\text{  when }0\le {\dot{h}_{n,k}}\le {\ddot{h}_{1}}, \\
 & \dddot{h}_{n,k,2},\text{  when }{\ddot{h}_{1}}< {\dot{h}_{n,k}}\le {\ddot{h}_{2}}, \\
 & \ \ \vdots \ \ \ \ \ \ \ \ \ \ \vdots  \\
 & \dddot{h}_{n,k,J},\text{   when }{\dot{h}_{n,k}}> {\ddot{h}_{J-1}}, \\
\end{split} \right.
\end{equation}
where the distance between IMD $k$ and BS $n$ is denoted by ${\dot{h}_{n,k}}$; $J-1$ elements in $\{ {\ddot{h}_{1}},{\ddot{h}_{2}},\cdots ,{\ddot{h}_{J-1}} \}$ are the distance thresholds; $j\text{-th}$ path component $\dddot{h}_{n,k,j}$ can be determined using
\begin{equation}\label{eq2}
\dddot{h}_{n,k,j}=\left\{ \begin{split}
  & {h}_{j}^\text{LS}\dot{h}_{n,k}^{-\gamma _{j}^\text{LS}},\text{  with }{\mathscr{P}_{j}}\left( \text{LoS}|{\dot{h}_{n,k}} \right), \\
 & {h}_{j}^{\text{NLS}}\dot{h}_{n,k}^{-\gamma _{j}^{\text{NLS}}},\text{  with }1-{\mathscr{P}_{j}}\left( \text{LoS}|{\dot{h}_{n,k}} \right). \\
\end{split} \right.
\end{equation}
\par
In equation \eqref{eq2}, the pathlosses at the reference-distance ${\dot{h}_{n,k}}=1$ for the LoS and NLoS cases are denoted by ${h}_{j}^\text{LS}$ and ${h}_{j}^{\text{NLS}}$ respectively, the pathloss exponents for the LoS and NLoS cases are represented by $\gamma _{j}^\text{LS}$ and $\gamma _{j}^{\text{NLS}}$ respectively, and a probability indicates the existence of the $j$-th LoS path component between IMD $k$ and BS $n$ at distance ${\dot{h}_{n,k}}$ is ${\mathscr{P}_{j}}( \text{LoS}|{\dot{h}_{n,k}} )$. Specifically, such a probability can be determined using
\begin{equation}\label{eq3}
{\mathscr{P}_{j}}\left\{ \text{LoS}|{\dot{h}_{n,k}} \right\}=\left\{ \begin{split}
  & 1-{{\dot{h}_{n,k}}}/{{{{\ddot{h}}}_{j}}}\;,\text{  }0<{\dot{h}_{n,k}}\le {{{\ddot{h}}}_{j}}, \\
 & \ \ \ \ 0,\ \ \ \ \ \ \ \ \ \ {\dot{h}_{n,k}}>{{{\ddot{h}}}_{j}}, \\
\end{split} \right.
\end{equation}
where ${{\ddot{h}}_{j}}$ is denoted as the distance threshold for $j\text{-th}$ path component.
\par
According to the resource utilization manner, we can easily know that only intra-cluster interferences exist for any IMD. Consequently, uplink data rate of IMD $k$ on subchannel $s$ of BS $n$ can be given by
\begin{equation}\label{eq4}
\left\{ \begin{aligned}
  & {{r}_{n,s,k}}=w{{\log }_{2}}\Big( 1+\frac{{{p}_{k}}{{h}_{n,k}}}{\sum\nolimits_{k\in {{\mathcal{Q}}_{n,s,k}}}{{{p}_{k}}{{h}_{n,k}}}+{{\sigma }^{2}}} \Big), \\
 & {{\mathcal{Q}}_{n,s,k}}=\left\{ k'\in \mathcal{K}_{-k} \right\}: {{h}_{n,k'}}\le {{h}_{n,k}},\\
 &\ \ \ \ \ \ \ \ \  \ \ {\dot{c}_{k}}={\dot{c}_{k'}}=s; {\dot{b}_{k}},{\dot{b}_{k'}}\in {{\dot{\mathcal{N}}}_{n}}, \\
\end{aligned} \right.
\end{equation}
where $\mathcal{K}_{-k}$ is the set $\mathcal{K}$ excluding $k$; ${\dot{c}_{k}}$ represents the subchannel index chosen by IMD $k$; ${\dot{b}_{k}}$ is the BS index chosen by IMD $k$; ${{\dot{\mathcal{N}}}_{n}}$ denotes the cluster that BS $n$ belongs to; $p_k$ is the transmission power of IMD $k$; $\sigma^2$ is the noise power.

\begin{figure}[!t]
\centering
\centerline{\includegraphics[width=3.5in]{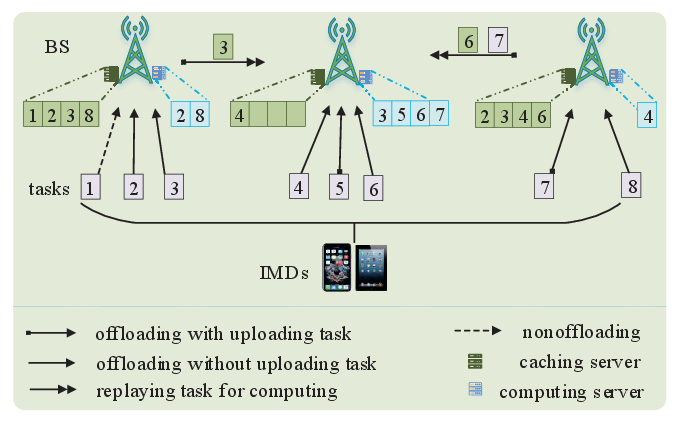}}
\caption{Caching and computing models.}
\label{fig2}
\end{figure}

\subsection{Caching and Computing Models}
Any task $m$ of IMD $k$ can be denoted as $\pi_{k,m}=\{d_{k,m}, {\ell}_{k,m}, {\tau}_{k,m}^{\max}, \rho_{k,m}\}$, where $d_{k,m}$ is the data size of task $k$; ${\ell}_{k,m}$ is the number of CPU cycles required for executing task $k$; ${\tau}_{k,m}^{\max}$ represents the deadline of task $k$; $\rho_{k,m}$ is the expected security level of task $k$.
\par
In this paper, a more complete version of the caching and computing models in \cite{TZhou2023Sen} are utilized. In addition, unlike the caching decisions in \cite{TZhou2023Sen}, we consider that tasks of IMDs are cached in advance according to their popularity. When IMD $k$ is associated with BS $n$, it initially verifies whether the BS has cached its tasks. If $\sum\nolimits_{n \in {\mathcal N}}{{u_{n,k,m}}}=0$, the task $m$ of IMD $k$ is computed locally, as illustrated by IMD 1 in Fig. \ref{fig2}, where $u_{n,k,m} \in \left\{ {0,1} \right\}$ represents the offloading (execution) indicator of task $m$ of IMD $k$ at BS $n$, with the constraint $\sum\nolimits_{n \in {\mathcal N}} {u_{n,k,m}}\le1$; ${u_{n,k,m}} = 1$ indicates that task $m$ of IMD $k$ is executed at BS $n$, while ${u_{n,k,m}} = 0$ otherwise. If ${x_{n,k}}{{c}_{n,k,m}}{u_{n,k,m}} = 1$, the task $m$ of IMD $k$ is directly computed at the selected BS $n$, and the results are downloaded to IMD $k$ from BS $n$, as exemplified by IMD 2 in Fig. \ref{fig2}. In this context, ${x_{n,k}} \in \left\{ {0,1} \right\}$ denotes the association indicator between IMD $k$ and BS $n$, with $\sum\nolimits_{n \in {\mathcal N}} {x_{n,k}}=1$; ${x_{n,k}} = 1$ signifies that IMD $k$ is associated with BS $n$, while ${x_{n,k}} =0$ otherwise. Furthermore, ${{c}_{n,k,m}} \in \left\{ {0,1} \right\}$ is the caching indicator of task $m$ at BS $n$, with $\sum\nolimits_{n \in {\mathcal N}} {{c}_{n,k,m}}\le1$; ${{c}_{n,k,m}} = 1$ implies that task $m$ is cached at BS $n$, while ${{c}_{n,k,m}} = 0$ otherwise. If ${x_{n,k}}{{c}_{n,k,m}}( {1 - {{c}_{n',k,m}}}){u_{n',k,m}} = 1$, the task $m$ of IMD $k$ is offloaded from the selected BS $n$ to another auxiliary BS $n' \ne n$ for computing via a wired link, as demonstrated by IMD 3 in Fig. \ref{fig2}. If ${x_{n,k}}{{c}_{n,k,m}}{{c}_{n',k,m}}{u_{n',k,m}} = 1$, the task $m$ of IMD $k$ is directly computed at auxiliary BS $n' \ne n$, as illustrated by IMD 4 in Fig. \ref{fig2}. When ${x_{n,k}}( {1 - {{c}_{n,k,m}}})( {1 - {{c}_{n',k,m}}}){u_{n,k,m}} = 1$, the task $m$ of IMD $k$ is offloaded to the selected BS $n$ for computing, as exemplified by IMD 5 in Fig. \ref{fig2}. In the case where ${x_{n,k}}( {1 - {{c}_{n,k,m}}}){{c}_{n',k,m}}{u_{n,k,m}} = 1$, the task $m$ of IMD $k$ is offloaded from another auxiliary BS $n' \ne n$ to the selected BS $n$ for computing via a wired link, as shown by IMD 6 in Fig. \ref{fig2}. If ${x_{n,k}}( {1 - {{c}_{n,k,m}}})( {1 - {{c}_{n',k,m}}}){u_{n',k,m}} = 1$, the task $m$ of IMD $k$ is initially offloaded to the selected BS $n$ and then transmitted to another auxiliary BS $n' \ne n$ for computing via a wired link, as demonstrated by IMD 7 in Fig. \ref{fig2}. Lastly, when ${x_{n,k}}( {1 - {{c}_{n,k,m}}}) {{c}_{n',k,m}}{u_{n',k,m}} = 1$, the task $m$ of IMD $k$ is directly calculated at an auxiliary BS $n' \ne n$, as illustrated by IMD 8 in Fig. \ref{fig2}.

\subsubsection{Local Computing} If $\sum\nolimits_{n \in {\mathcal N}} {{u_{n,k,m}}}=0$ is satisfied, the task $m$ of IMD $k$ is executed locally. Consequently, the local executing time (latency) of task $m$ of IMD $k$ is
\begin{equation}\label{eq5}
\tau _{k,m}^\text{locc}={{{\ell}_{k,m}}(1-\sum\nolimits_{n \in {\mathcal N}} {{u_{n,k,m}}})}/f_{k}^\text{loc},
\end{equation}
where $f_{k}^{\text{loc}}$ is the computation capacity of IMD $k$. Then, the local executing energy consumption of task $m$ of IMD $k$ is
\begin{equation}\label{eq6}
\begin{split}
\varepsilon _{k,m}^\text{locc}=\alpha {{\left( f_{k}^\text{loc} \right)}^{3}}\tau _{k,m}^\text{locc}\overset{(a)}{\mathop{=}}\alpha { {{\ell}_{k,m}}(1-\sum\limits_{n \in {\mathcal N}}{{u_{n,k,m}}})}{{\left( f_{k}^\text{loc} \right)}^{2}},\\
\end{split}
\end{equation}
where $\alpha$ is the energy coefficient of chip architecture; (a) follows from a fact that the task $m$ of IMD $k$ can be executed at most one BS.

\subsubsection{Edge Computing} If ${x_{n,k}}( {1 - {{c}_{n,k,m}}})( {1 - {{c}_{n',k,m}}}){u_{n,k,m}} \\ = 1$ or ${x_{n,k}}( {1 - {{c}_{n,k,m}}})( {1 - {{c}_{n',k,m}}}){u_{n',k,m}} = 1$ is satisfied under $n' \ne n$, the task $m$ of IMD $k$ needs to be uploaded to BS $n'$ or $n$ for computing through a NOMA manner, respectively. Consequently, the uploading time of task $m$ of IMD $k$ is
\begin{equation}\label{eq7}
\begin{split}
  & \tau _{k,m}^{\text{up}}=\sum\nolimits_{n\in \mathcal{N}}\sum\nolimits_{s\in \mathcal{S}}{{{z}_{s,k}}{{d}_{k,m}}}/{{{r}_{n,s,k}}}(\\
  & \quad {{x}_{n,k}}(1-{{c}_{n,k,m}}){{u}_{n,k,m}}\sum\nolimits_{{n}'\in \mathcal{N}_{-n}}(1-{{c}_{{n}',k,m}}) \\
 & \quad +{{x}_{n,k}}(1-{{c}_{n,k,m}})\sum\nolimits_{{n}'\in \mathcal{N}_{-n}}{(1-{{c}_{{n}',k,m}}){{u}_{{n}',k,m}}} ), \\
\end{split}
\end{equation}
where ${z}_{s,k} \in\{0,1\}$; ${z}_{s,k}=1$ if IMD $k$ selects the subchannel $s$, 0 otherwise. Then, the uploading energy consumption of task $m$ of IMD $k$ is
\begin{equation}\label{eq8}
\begin{split}
   & \varepsilon _{k,m}^{\text{up}}=\sum\nolimits_{n\in \mathcal{N}}\sum\nolimits_{s\in \mathcal{S}}{{{z}_{s,k}}{{p}_{k}}{{d}_{k,m}}}/{{{r}_{n,s,k}}}(\\
  & \quad {{x}_{n,k}}(1-{{c}_{n,k,m}}){{u}_{n,k,m}}\sum\nolimits_{{n}'\in \mathcal{N}_{-n}}(1-{{c}_{{n}',k,m}}) \\
 & \quad +{{x}_{n,k}}(1-{{c}_{n,k,m}})\sum\nolimits_{{n}'\in \mathcal{N}_{-n}}{(1-{{c}_{{n}',k,m}}){{u}_{{n}',k,m}}} ), \
\end{split}
\end{equation}
 In addition, if ${x_{n,k}}{{c}_{n,k,m}}( {1 - {{c}_{n',k,m}}}){u_{n',k,m}} = 1$ or ${x_{n,k}}( {1 - {{c}_{n,k,m}}})( {1 - {{c}_{n',k,m}}}){u_{n',k,m}} = 1$ is satisfied under $n' \ne n$, the task $m$ of IMD $k$ needs to be transmitted from the  BS $n$ selected by this IMD to an auxiliary BS $n'$ through a wired link; if ${x_{n,k}}( {1 - {{c}_{n,k,m}}}){{c}_{n',k,m}}{u_{n,k,m}} = 1$, the task $m$ of IMD $k$ is offloaded from another auxiliary BS $n' \ne n$ to the BS $n$ selected by IMD $k$ for computing through a wired link. Then, the corresponding backhaul time of task $m$ of IMD $k$ is
\begin{equation}\label{eq9}
\begin{split}
  & \tau _{k,m}^{\text{bh}}=\sum\nolimits_{n\in \mathcal{N}}{{{{{d}_{k,m}}}/{{{r}^{\text{bh}}}}}}\left(\right.\\
  &\quad \ \ {{x}_{n,k}}{{c}_{n,k,m}}\sum\nolimits_{n'\in \mathcal{N}_{-n}}{(1-{{c}_{{n}',k,m}}){{u}_{{n}',k,m}}}\\
&\quad +{{x}_{n,k}}(1-{{c}_{n,k,m}}){{u}_{{n},k,m}}\sum\nolimits_{n'\in \mathcal{N}_{-n}}{{{c}_{{n}',k,m}}}\\
 &\quad  \left. +{{x}_{n,k}}(1-{{c}_{n,k,m}})\sum\nolimits_{n'\in \mathcal{N}_{-n}}{(1-{{c}_{{n}',k,m}}){{u}_{{n}',k,m}}} \right), \\
&=\sum\nolimits_{n\in \mathcal{N}}{{\frac{{{d}_{k,m}}}{{{r}^{\text{bh}}}}}}\left({{x}_{n,k}}\sum\nolimits_{n'\in \mathcal{N}_{-n}}{(1-{{c}_{{n}',k,m}}){{u}_{{n}',k,m}}}  \right. \\
 &\quad  \left. +{{x}_{n,k}}(1-{{c}_{n,k,m}}){{u}_{{n},k,m}}\sum\nolimits_{n'\in \mathcal{N}_{-n}}{{{c}_{{n}',k,m}}} \right), \\
\end{split}
\end{equation}
where ${r^\text{bh}}$ represents the backhual rate between two BSs. The remote executing time of task $m$ of IMD $k$ is
\begin{equation}\label{eq10}
\begin{aligned}
 & \tau _{k,m}^{\text{mecc}}=\sum\nolimits_{n\in \mathcal{N}}\left( {{x}_{n,k}}(1-{{c}_{n,k,m}}){{u}_{n,k,m}}\dot{\tau}_{n,k,m}^\text{mec} \right. \\
 & +{{x}_{n,k}}{{c}_{n,k,m}}{{u}_{n,k,m}}\dot{\tau}_{n,k,m}^\text{mec} \\
 & +{{x}_{n,k}}(1-{{c}_{n,k,m}})\sum\nolimits_{n'\in \mathcal{N}_{-n}}{{{c}_{{n}',k,m}}{{u}_{{n}',k,m}}}\dot{\tau}_{n',k,m}^\text{mec} \\
 &+{{x}_{n,k}}(1-{{c}_{n,k,m}})\sum\nolimits_{n'\in \mathcal{N}_{-n}}{(1-{{c}_{{n}',k,m}}){{u}_{{n}',k,m}}}\dot{\tau}_{n',k,m}^\text{mec} \\
 &+{{x}_{n,k}}{{c}_{n,k,m}}\sum\nolimits_{n'\in \mathcal{N}_{-n}}{{{c}_{{n}',k,m}}{{u}_{{n}',k,m}}}\dot{\tau}_{n',k,m}^\text{mec} \\
 &\left. +{{x}_{n,k}}{{c}_{n,k,m}}\sum\nolimits_{n'\in \mathcal{N}_{-n}}{(1-{{c}_{{n}',k,m}}){{u}_{{n}',k,m}}}\dot{\tau}_{n',k,m}^\text{mec}\right) \\
&=\sum\nolimits_{n\in \mathcal{N}}{{x}_{n,k}}\sum\nolimits_{n'\in \mathcal{N}}{{{u}_{{n}',k,m}}}\dot{\tau}_{n',k,m}^\text{mec}\overset{(b)}{\mathop{=}}\sum\nolimits_{n\in \mathcal{N}}{{{u}_{{n},k,m}}}\dot{\tau}_{n,k,m}^\text{mec}, \\
\end{aligned}
\end{equation}
where (b) follows from a fact that IMD $k$ can be associated with only one BS; $\dot{\tau}_{n,k,m}^\text{mec}={{{{\ell}_{k,m}}}/{f_{n,k,m}^{\text{mec}}}}$ is the executing time of task $m$ of IMD $k$ at BS $n$. Based on the ratio of the CPU cycles required for processing task $m$ of IMD $k$ to the CPU cycles available at BS $n$, the computing capacity (capability) of BS $n$ is allocated for the computation and secure operations of this task. In particular, when the task $m$ of IMD $k$ is executed at BS $n$, the computing capacity $f_{n,k,m}^\text{mecc}$ assigned to this task by BS $n$ can be given by
\begin{equation}\label{eq11}
\left\{ \begin{aligned}
  & f_{n,k,m}^{\text{mec}}=\frac{f_n^{\text{mmax}}\mathscr{A}_{k,m,l}}{\sum\nolimits_{k'\in \mathcal{K}}{\sum\nolimits_{m'\in \mathcal{M}}{{{u}_{n,k',m'}}\mathscr{A}_{k',m',l}}}}, \\
 & \mathscr{A}_{k,m,l}={{\ell}_{k,m}}+\sum\nolimits_{l\in \mathcal{L}}{{{v}_{k,m,l}}}{{{\dot{\epsilon}}}_{l}}{{d}_{k,m}}, \\
\end{aligned} \right.
\end{equation}
where $f_n^{\text{mmax}}$ is the maximum computing capacity of BS $n$.

\subsection{Security Model}
In reality, computation tasks often have different security requirements, and they are vulnerable to malicious attacks, spoofing, and eavesdropping when these tasks are offloaded to edge servers. To tackle this issue, various cryptographic algorithms are used for data encryption and decryption in MEC networks. As revealed in \cite{MZahed2020}, the energy consumption and latency significantly increase with the robustness and strength of these algorithms. In addition, these precautions cannot prevent security breaches absolutely. Therefore, the quantification of security risks is a significant challenge for secure and green computation offloading.
\par
As we know, different cryptographic algorithms correspond to different security levels. By employing these algorithms, the offloaded tasks are encrypted and decrypted on IMDs and edge servers, respectively. It is evident that such operations are computationally expensive and cause additional overhead (e.g., latency and energy). It is assumed that there are $L$ cryptographic algorithms indexed within the set $\mathcal{L}=\{1,2,\cdots,L\}$, the protection level (robustness) of any cryptographic algorithm $l$ is ${\dot{\rho}}_l=l$, and a higher ${\dot{\rho}}_l$ means stronger robustness. In addition, for cryptographic algorithm $l$, its computation capacities used for encrypting and decrypting one bit are $\epsilon_l$ (in CPU cycles/bit) and $\dot{\epsilon}_l$ (in CPU cycles/bit) respectively, and its energy consumptions used for encrypting and decrypting one bit are assumed to be the same, i.e., $\ddot{\epsilon}_l$ (in mJ /bit). We let $\boldsymbol{\epsilon}=\{\epsilon_l,\forall l \in \mathcal{L}\}$, $\boldsymbol{\dot{\epsilon}}=\{\dot{\epsilon}_l,\forall l \in \mathcal{L}\}$ and $\boldsymbol{\ddot{\epsilon}}=\{\ddot{\epsilon}_l,\forall l \in \mathcal{L}\}$.
\par
In this paper, we assume that the tasks cached at BSs have been encrypted. When some task needs to be uploaded from an IMD to a BS for computing, this IMD encrypts it before transmission. Whether some task has been cached at a BS or received from an IMD or another BS, this BS can execute it after decryption. Evidently, the encrypting operations are done by IMDs, and the decrypting operations are performed by BSs that execute (calculate) tasks. When cryptographic algorithm $l$ is employed for task $m$ of IMD $k$, the failure probability \cite{WJiang2015Aug} can be given by
\begin{equation}\label{eq12}
\dot{\mathscr{P}}_{k,m,l}=\left\{\begin{array}{l}
		1-e^{-{\theta}_{k,m}\left(\rho_{k,m}-{\dot{\rho}}_l\right)}, \text { if } {\dot{\rho}}_l<\rho_{k,m}, \\
		0, \text { otherwise},
	\end{array}\right.
\end{equation}
where ${\theta}_{k,m}$ denotes the security risk coefficient associated with task $m$ of IMD $k$. As shown in \eqref{eq12}, algorithm $l$ effectively safeguards task $m$ of IMD $k$ from attacks if its security level is at least as high as the level required by the task. Conversely, if the security level of algorithm algorithm $l$ is inadequate, it fails to protect task $m$ of IMD $k$ with a certain probability.
\par
Local encrypting time of task $m$ of IMD $k$ can be given by
\begin{equation}\label{eq13}
\tau _{k,m}^{\text{loce}}=\sum\nolimits_{n\in \mathcal{N}}{\sum\nolimits_{l\in \mathcal{L}}{{{u}_{n,k,m}}{{v}_{k,m,l}}}{{{{\epsilon }}}_{l}}{{d}_{k,m}}/f_{k}^\text{loc}},
\end{equation}
and its local encrypting energy consumption is
\begin{equation}\label{eq14}
\varepsilon _{k,m}^\text{loce}=\sum\nolimits_{n\in \mathcal{N}}{\sum\nolimits_{l\in \mathcal{L}}{{{u}_{n,k,m}}{{v}_{k,m,l}}}{{{\ddot{\epsilon}}}_{l}}{{d}_{k,m}}},
\end{equation}
where ${v}_{k,m,l}$ is a decision variable; ${v}_{k,m,l}=1$ if the task $m$ of IMD $k$ selects cryptographic algorithm $l$, 0 otherwise.
\par
Then, the remote decrypting time of task $m$ of IMD $k$ is
\begin{equation}\label{eq15}
\tau _{k,m}^{\text{mecd}}=\sum\nolimits_{n\in \mathcal{N}}{\sum\nolimits_{l\in \mathcal{L}}{{{u}_{n,k,m}}{{v}_{k,m,l}}}{{{\dot{\epsilon}}}_{l}}{{d}_{k,m}}/f_{n,k,m}^\text{mec}}.
\end{equation}
\par
The security breach cost \cite{WJiang2015Aug} of task $m$ of IMD $k$ is
\begin{equation}\label{eq16}
\varphi_{k,m}=\sum\nolimits_{n\in \mathcal{N}}{\sum\nolimits_{l \in \mathcal{L}} {\lambda_{k,m}}{{u}_{n,k,m}}{{v}_{k,m,l}} \dot{\mathscr{P}}_{k,m,l}},
\end{equation}
where $\lambda_{k,m}$ represents the financial loss incurred if task $m$ of IMD $k$ fails, with the loss measured in United States dollars (USD). Consequently, the total cost of security breaches for IMD $k$ can be given by $\psi_k=\sum\nolimits_{m\in \mathcal{M}} \varphi_{k,m}$.

\section{Problem Formulation}\label{sec3}
According to descriptions in previous section, the processing delay (time) of task $m$ of IMD $k$ can be given by
\begin{equation}\label{eq17}
{{\tau }_{k,m}}=\tau _{k,m}^\text{locc}+\tau _{k,m}^\text{loce}+\tau _{k,m}^\text{up}+\tau _{k,m}^{\text{bh}}+\tau _{k,m}^\text{mecd}+\tau _{k,m}^\text{mecc},
\end{equation}
and the local processing energy consumption of all tasks of IMD $k$ can be given by
\begin{equation}\label{eq18}
\varepsilon_k^\text{loc}=\sum\nolimits_{m\in \mathcal{M}}\left(\varepsilon_{k,m}^\text{locc}+\varepsilon_{k,m}^\text{loce}+\varepsilon_{k,m}^\text{up}\right).
\end{equation}
\par
To achieve green and secure communications, we concurrently carry out device associations, caching, computation offloading, channel selections, power controls, security service assignments, and local computing capacity adjustments. Our objective is to minimize the energy consumption of IMDs while adhering to constraints related to task processing time, uplink power, local computing capacity, caching capacity, and security costs. Mathematically, it is
\begin{equation}\label{eq19}
\begin{aligned}
 &\underset{\mathbf{\Xi}}{\mathop{\min}}\,E\left(\mathbf{\Xi}\right)=\sum\nolimits_{k \in \mathcal{K}} \varepsilon_k^\text{loc} \\
&\ \ \text{s.t. }{C_1: }{{\tau }_{k,m}}\le \tau _{k,m}^{\max },\forall k\in \mathcal{K},\forall m\in \mathcal{M},\\
&\ \ \ \ \ {C_2: }\sum\nolimits_{n\in \mathcal{N}}{{{x}_{n,k}}}= 1,\forall k\in \mathcal{K},\\
&\ \ \ \ \ {C_3: }\sum\nolimits_{s\in \mathcal{S}}{{{z}_{s,k}}}=1,\forall k\in \mathcal{K},\\
&\ \ \ \ \ {C_4: }\sum\nolimits_{n\in \mathcal{N}}{{{u}_{n,k,m}}}\le 1,\forall k\in \mathcal{K},\forall m\in \mathcal{M},\\
&\ \ \ \ \ {C_5: }\sum\nolimits_{l\in \mathcal{L}}{{{v}_{k,m,l}}}=1,\forall k\in \mathcal{K},\forall m\in \mathcal{M},\\
 & \ \ \ \ \ {C_6: }\psi_{k}\le \psi_{k}^{\max},\forall k\in \mathcal{K}, \\
 & \ \ \ \ \ {C_7: }\vartheta \le f_{k}^\text{loc}\le {f}_{k}^{\text{lmax}},\forall k\in \mathcal{K},\\
 &\ \ \ \ \ {C_8: }\vartheta \le {{p}_{k}}\le p_{k}^{\max },\forall k\in \mathcal{K},\\
 &\ \ \ \ \ {C_9: }{{x}_{n,k}}\in \left\{ 0,1 \right\},\forall n\in \mathcal{N},\forall k\in \mathcal{K},\\
 & \ \ \ \ \ {C_{10}: }{{z}_{s,k}}\in \left\{ 0,1 \right\},\forall s\in \mathcal{S},\forall k\in \mathcal{K},\\
 &\ \ \ \ \ {C_{11}: }{{u}_{n,k,m}}\in \left\{ 0,1 \right\},\forall n\in \mathcal{N},\forall k\in \mathcal{K},\forall m\in \mathcal{M},\\
 &\ \ \ \ \ {C_{12}: }{{v}_{k,m,l}}\in \left\{ 0,1 \right\},\forall k\in \mathcal{K},\forall m\in \mathcal{M}, \forall l\in \mathcal{L},\\
\end{aligned}
\end{equation}
where $\mathbf{\Xi}=\{\mathbf{X},\mathbf{Z},\mathbf{U},\mathbf{V},{{\mathbf{f}}^\text{loc}},\mathbf{p}\}$; $\mathbf{X}=\{ {{x}_{n,k}},\forall n\in \mathcal{N},\forall k\in \mathcal{K} \}$; $\mathbf{Z}=\{ {{z}_{s,k}},\forall s\in \mathcal{S},\forall k\in \mathcal{K}\}$; $\mathbf{U}=\{ {{u}_{n,k,m}},\forall n\in \mathcal{N},\forall k\in \mathcal{K},\forall m\in \mathcal{M} \}$; $\mathbf{V}=\{ {{v}_{k,m,l}},\forall k\in \mathcal{K},\forall m\in \mathcal{M},\forall l\in \mathcal{L}\}$;  ${{\mathbf{f}}^\text{loc}}=\{ f_{k}^\text{loc},\forall k\in \mathcal{K} \}$; $\mathbf{p}=\{ {{p}_{k}},\forall k\in \mathcal{K} \}$; the constraint $C_1$ shows the processing time of task $m$ of IMD $k$ cannot exceed its deadline $\tau _{k,m}^{\max }$; $C_2$ and $C_9$ reveal that IMD $k$ can only be associated with a single BS;  $C_3$ and $C_{10}$ mean that IMD $k$ can only utilize one subchannel; $C_4$ and $C_{11}$ specify that the task $m$ of IMD $k$ can be executed on at most one BS; $C_5$ and $C_{12}$ show that the task $m$ of IMD $k$ can only select one cryptographic algorithm; $C_6$ stipulates that overall security breach cost of IMD $k$ must not surpass its maximum permissible cost, denoted as $\psi_{k}^{\max}$; $C_7$ gives the minimum and maximum allowable computation capacities for IMD $k$, ${f}_{k}^{\text{lmax}}$ represents its maximum allowable capacity, and $\vartheta $ is set to a sufficiently small value (e.g., $10^{-20}$) to avoid zero division; $C_5$ stipulates lower and upper limits of transmission power for IMD $k$, and $p_{k}^{\max }$ is its maximum allowable power.

\section{Algorithm design}\label{sec4}
Problem \eqref{eq19} takes a mixed-integer form, wherein its optimization variables are coupling. This means that some conventional optimization methods may not be suitable for tackling such a problem. To solve it, we further improve the existing algorithm (i.e., IHAS \cite{TZhou2022Oct}), and develop FIHAS. In such an algorithm, IADGGA is first used for finding a coarse-grained solution, and then adaptive PSO (APSO) is utilized to search for a fine-grained solution. Unlike the ADGGA in HAS, we consider new highly effective crossover and mutation probabilities in IADGGA to search the feasible space of the formulated problem more fully, which can further enhance the population diversity and avoid premature convergence. In addition, unlike ADGGA, at the latter stage of iteration of IADGGA, the individuals with high similarity and low fitness are eliminated and then regenerated using the rules of population initialization. Such an operation can further enhance the population diversity to achieve better solutions.

\subsection{IADGGA}
To solve the problem \eqref{eq19} using GA, a series of genetic operations need to be done repeatedly until an effective solution is found or the iterative process converges. Such operations involve genetic selection, crossover and mutation. The first serves to inherit good patterns from the current population to the next generation, whereas the latter two are instrumental in maintaining population diversity. The key factors and corresponding steps of GA are as follows.

\subsubsection{Chromosomes} In GA, the population comprises numerous individuals, each constituted by distinct chromosomes. Significantly, any individual is a solution to the optimization problem, and individuals are indexed from 1 to $I$ within $\mathcal{I}=\{1,2,\cdots,I\}$. To solve problem \eqref{eq19} using GA, parameters $\mathbf{X}$, $\mathbf{U}$, $\mathbf{Z}$, $\mathbf{U}$, $\mathbf{V}$, ${{\mathbf{f}}^\text{loc}}$, and $\mathbf{p}$ are encoded as $\mathbf{\bar{X}}_{i}=\{\bar{x}_{i,k}, \forall k \in \mathcal{K}\}$  $\mathbf{\bar{Z}}_{i}=\{\bar{z}_{i,k}, \forall k \in \mathcal{K}\} $, ${\mathbf{\bar{U}}}_{i}=\{{\bar{u}}_{i,k}, \forall k \in \dot{\mathcal{K}}\}$, ${\mathbf{\bar{V}}}_{i}=\{{\bar{v}}_{i,k}, \forall k \in \dot{\mathcal{K}}\}$, $\mathbf{\bar{F}}_{i}=\{{\bar{f}}_{i,k}, \forall k \in \mathcal{K}\}$,  and $\mathbf{\bar{P}}_{i}=\{{\bar{p}}_{i,k}, \forall k \in \mathcal{K}\}$ for any individual ${i}\in \mathcal{I}$ respectively. In this context, all tasks associated with the $K$ IMDs are treated as virtual IMDs, which are indexed from 1 to $KM$ within $\dot{\mathcal{K}}=\{1,2, \cdots, KM\} $. In individual $i$, $\bar{x}_{i,k}$ signifies the BS index chosen by IMD $k$, $\bar{z}_{i,k}$ indicates the subchannel index chosen by IMD $k$, ${\bar{u}}_{i,k}$ represents the BS index that serves virtual IMD $k$, ${\bar{v}}_{i,k}$ is the cryptographic algorithm index chosen by virtual IMD $k$, ${\bar{f}}_{i,k}$ stands for the computing capacity of IMD $k$, and ${\bar{p}}_{i,k}$ refers to the transmission power of IMD $k$. In addition, $\mathbf{\bar{X}}=\{\bar{x}_{i,k}, \forall i \in \mathcal{I}, \forall k \in {\mathcal{K}}\} $,  $\mathbf{\bar{Z}}=\{\bar{z}_{i,k}, \forall i \in \mathcal{I}, \forall k \in \mathcal{K}\} $, ${\mathbf{\bar{U}}}=\{{\bar{u}}_{i,k}, \forall i \in \mathcal{I}, \forall k \in \dot{\mathcal{K}}\}$, ${\mathbf{\bar{V}}}=\{{\bar{v}}_{i,k}, \forall i \in \mathcal{I}, \forall k \in \dot{\mathcal{K}}\}$, $\mathbf{\bar{F}}=\{{\bar{f}}_{i,k}, \forall i \in \mathcal{I}, \forall k \in \mathcal{K}\}$, and $\mathbf{\bar{P}}=\{{\bar{p}}_{i,k}, \forall i \in \mathcal{I}, \forall k \in \mathcal{K}\}$.

\subsubsection{Fitness function} In GA, fitness functions are mainly used to assess the fitness of individuals. Through direct observation, we can easily find that constraints $C_1$ and $C_6$ of problem \eqref{eq19} are in a mixed-integer coupling form, they may not be met in genetic operations and initialization of GA well. Based on this, these constraints are incorporated into the fitness function as penalty components. As a result, the fitness function for any individual ${i}$ can be given by
\begin{equation}\label{eq20}
	\begin{aligned}
		&G(\mathbf{\bar{\Xi}}_{i})=-E(\mathbf{\bar{\Xi}}_{i})-\sum\nolimits_{k \in \mathcal{K}} {\eta}_k \max \left(\psi_{k}-\psi_{k}^{\max}, 0\right)\\
           &\ \ \ \ \ \ \ \ \ -\sum\nolimits_{k \in \mathcal{K}} \sum\nolimits_{m\in \mathcal{M}} \tilde{\eta}_{k,m} \max \left(\tau_{k,m}-\tau_{k,m}^{\max }, 0\right),\\
	\end{aligned}
\end{equation}
where $\mathbf{\bar{\Xi}}_{i}=\{\mathbf{\bar{X}}_{i}, \mathbf{\bar{U}}_{i}, \mathbf{\bar{Z}}_{i}, {\mathbf{\bar{U}}}_{i},  {\mathbf{\bar{V}}}_{i}, \mathbf{\bar{F}}_{i},\mathbf{\bar{P}}_{i}\}$; ${\eta}_k$ is the penalty factor of IMD $k$; $\tilde{\eta}_{k,m}$ is the penalty factor of task $m$ of IMD $k$.

\subsubsection{Population initialization} To meet constraints other than $C_1$ and $C_6$ in problem \eqref{eq19}, the initial population is generated based on the following guidelines. Particularly, the initial genes of any individual ${i}$ are determined by
\begin{equation}\label{eq21}
\left\{ \begin{aligned}
  & \bar{x}_{i,k}^{0}=\dot{R}(\mathcal{N}),\forall k\in {\mathcal{K}}, \\
 & \bar{z}_{i,k}^{0}=\dot{R}(\mathcal{S}),\forall k\in \mathcal{K}, \\
 & \bar{u}_{i,k}^{0}=\dot{R}(\mathcal{N}\cup \left\{ 0 \right\}),\forall k\in \dot{\mathcal{K}}, \\
 & \bar{v}_{i,k}^{0}=\dot{R}(\mathcal{L}),\forall k\in \dot{\mathcal{K}}, \\
 & \bar{f}_{i,k}^{0}=\ddot{R}(f_{k}^{\text{lmax}}),\forall k\in \mathcal{K}, \\
 & \bar{p}_{i,k}^{0}=\ddot{R}(p_{k}^{\max }),\forall k\in \mathcal{K}, \\
\end{aligned} \right.
\end{equation}
where $\dot{R}(\mathcal{B})$ outputs an element from $\mathcal{B}$ randomly; $\ddot{R}(b)$ generates a value from the interval $(0,b)$ randomly; $\bar{u}_{i,k}^{0}=0$ means that the task of an IMD corresponding to virtual IMD $k$ executed locally.

\subsubsection{Selection} In this study, the tournament selection method is employed for choosing individuals in the GA. To further boost its effectiveness, the historically best individual is consistently retained within the population. This means that if the historically best individual is not chosen for the subsequent generation, the worst selected individual should be replaced by the historically best one. During each iteration, the historically best individual must be updated accordingly. Notably, the best individual is identified as the one with the highest fitness value among all individuals, while the worst is the opposite; the current best individual signifies the best within the current generation, and the historically best individual is the top individual across all generations.

\subsubsection{Crossover} 
\begin{figure}[!t]
\centering
\centerline{\includegraphics[width=3.8in]{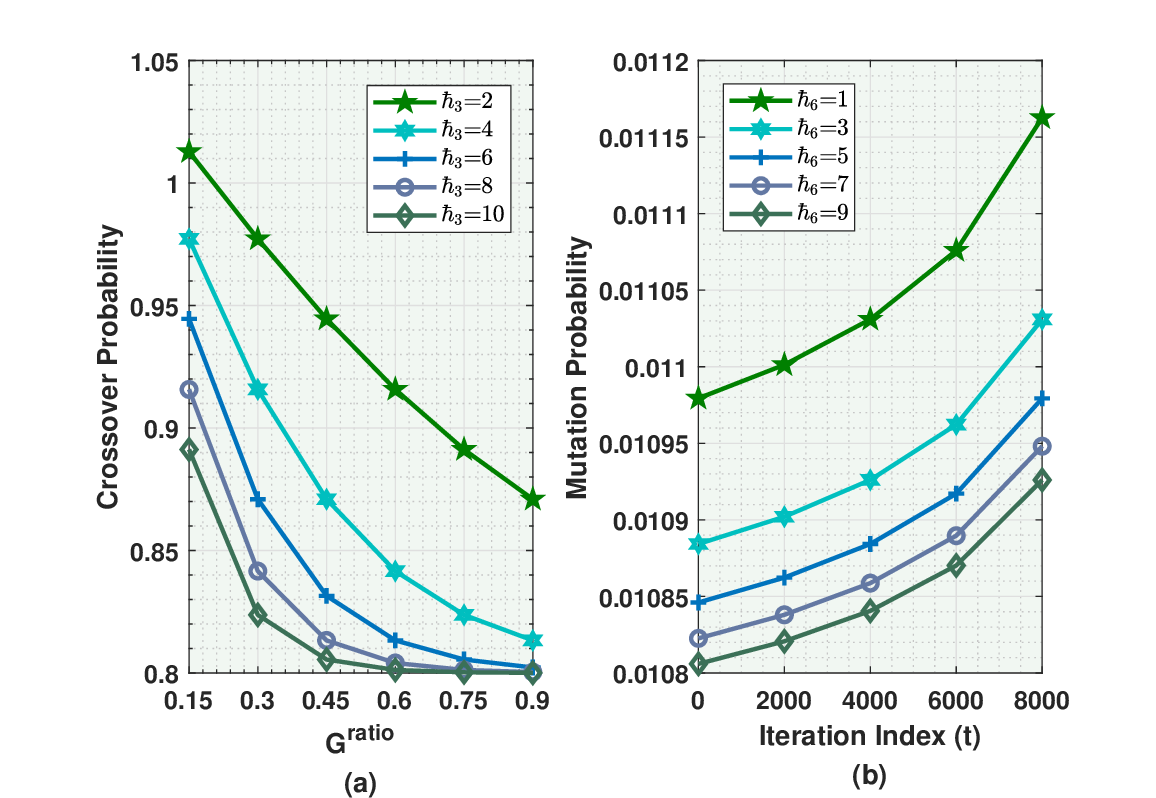}}
\caption{Crossover and mutation probabilities.}
\label{fig3}
\end{figure}

In this paper, any two adjacent paternal individuals ${i}$ and $i'=i+1$ are selected to perform crossover operations on genes with probability $\ddot{\mathscr{P}}_{i,i'}$. After crossover, two offsprings (new individuals) are generated. The crossover probability $\ddot{\mathscr{P}}_{i,i'}$ between individuals ${i}$ and $i'=i+1$ \cite{YZhou2022} is
\begin{equation}\label{eq22}
\ddot{\mathscr{P}}_{i,i'}=\left\{ \begin{split}
  & {{\hbar}_{1}}+\frac{{{\hbar}_{1}}-{{\hbar}_{2}}}{1+{{e}^{{{\hbar}_{3}}G^\text{ratio}}}}, {{{\dot{G}}}_{i,i'}}\ge {{G}^\text{ave}}, \\
 & {{\hbar}_{1}}, \ \ \ \ \ \ \ \ \ \ \ {{{\dot{G}}}_{i,i'}}<{{G}^\text{ave}}, \\
\end{split} \right.
\end{equation}
where $e\approx 2.7183$; ${\hbar}_{1}$ and ${\hbar}_{2}$ are the maximum and minimum crossover probabilities, respectively; ${\hbar}_{3}$ is the curve smoothing parameter; $\dot{G}_{i,i'}$ is the maximum of fitness (function) values of individuals ${i}$ and $i'=i+1$; $G^{\text{ave}}$ and $G^{\max }$ represent the average and maximum fitness values of the population, respectively; $G^\text{ratio}={( {{{\dot{G}}}_{i,i'}}-{{G}^\text{ave}} )}/{( {{G}^{\max }}-{{G}^\text{ave}} )}$. Fig.\ref{fig3} (a) shows the crossover probability $\ddot{\mathscr{P}}_{i,i'}$ between individuals ${i}$ and $i'=i+1$ under ${{{\dot{G}}}_{i,i'}}\ge {{G}^\text{ave}}$. As illustrated in Fig.\ref{fig3} (a), ${\hbar}_{3}$ is used for adjusting the gradient of the crossover probability curve. When $\dot{G}_{i,i'}$ approaches ${{G}^{\max }}$, $\ddot{\mathscr{P}}_{i,i'}$ decreases to maintain good individuals; when $\dot{G}_{i,i'}$ nears ${{G}^\text{ave}}$, $\ddot{\mathscr{P}}_{i,i'}$ increases to generate better offspring and thus enhance the population diversity. Compared with the linear crossover probability of ADGGA in \cite{TZhou2022Oct}, the crossover probability \eqref{eq22} that will be utilized in IADGGA lets individuals have more opportunities to maintain good individuals when $\dot{G}_{i,i'}$ nears ${{G}^{\max }}$, and generate better offspring when $\dot{G}_{i,i'}$ nears ${{G}^\text{ave}}$.

\subsubsection{Mutation} In the mutation operation, any individual ${i}$ is first selected randomly, and its genes are changed with probability $\dddot{\mathscr{P}}_{i}$ under the constraints of problem \eqref{eq19}. The mutation probability \cite{YZhou2022} of individual ${i}$ can be given by
\begin{equation}\label{eq23}
\dddot{\mathscr{P}}_{i}={{\hbar}_{4}}+\frac{{{\hbar}_{5}}}{1+\ln \left[ {{\hbar}_{6}}\left( {{T}_{1}}-{t}+1 \right) \right]},
\end{equation}
where ${\hbar}_{4}$, ${\hbar}_{5}$, and ${\hbar}_{6}$ are positive real numbers; $t$ and ${T}_{1}$ are the iteration index and the number of iterations, respectively; $\ln(b)$ is a logarithmic function with base $e$ in terms of parameter $b$. As illustrated in Fig.\ref{fig3} (b), ${\hbar}_{6}$ is used for adjusting the gradient of the mutation probability curve. The mutation probability of any individual increases with the iteration index to enhance the search capability of the latter stages. Compared with the linear mutation probability of ADGGA in \cite{TZhou2022Oct}, the mutation probability \eqref{eq23} that will be utilized in IADGGA lets individuals have more opportunities to mutate in the latter stages, which means a higher probability of avoiding premature convergence.
\par
Then, the genes of any individual ${i}$ can perform mutation operations with probability $\dddot{\mathscr{P}}_{i}$ using
\par
\begin{equation}\label{eq24}
	\bar{x}_{i,k}=\left\{\begin{aligned}
		&R\left({{\kappa}_{1}}N+\left(1-{{\kappa}_{1}}\right) \bar{x}_{i,k}\right),{{\kappa}_{2}}>0.5, \forall k \in {\mathcal{K}}, \\
		&R\left({{\kappa}_{1}}+\left(1-{{\kappa}_{1}}\right) \bar{x}_{i,k}\right), {{\kappa}_{2}} \leq 0.5, \forall k \in {\mathcal{K}},
	\end{aligned}\right.
\end{equation}
\begin{equation}\label{eq25}
	\bar{z}_{i,k}=\left\{\begin{aligned}
		&R\left({{\kappa}_{1}}S+\left(1-{{\kappa}_{1}}\right) \bar{z}_{i,k}\right), {{\kappa}_{2}}>0.5, \forall k \in \mathcal{K}, \\
		&R\left({{\kappa}_{1}}+\left(1-{{\kappa}_{1}}\right) \bar{z}_{i,k}\right), {{\kappa}_{2}} \leq 0.5, \forall k \in \mathcal{K},
	\end{aligned}\right.
\end{equation}
\begin{equation}\label{eq26}
	\bar{u}_{i,k}=\left\{\begin{aligned}
		&R\left({{\kappa}_{1}}N+\left(1-{{\kappa}_{1}}\right) \bar{u}_{i,k}\right),{{\kappa}_{2}}>0.5, \forall k \in \dot{\mathcal{K}}, \\
		&R\left(\left(1-{{\kappa}_{1}}\right) \bar{u}_{i,k}\right), {{\kappa}_{2}} \leq 0.5, \forall k \in \dot{\mathcal{K}},
	\end{aligned}\right.
\end{equation}
\begin{equation}\label{eq27}
	\bar{v}_{i,k}=\left\{\begin{aligned}
		&R\left({{\kappa}_{1}} L+\left(1-{{\kappa}_{1}}\right) \bar{v}_{i,k}\right), {{\kappa}_{2}}>0.5, \forall k \in \dot{\mathcal{K}}, \\
		&R\left({{\kappa}_{1}}+\left(1-{{\kappa}_{1}}\right) \bar{v}_{i,k}\right), {{\kappa}_{2}} \leq 0.5, \forall k \in \dot{\mathcal{K}},
	\end{aligned}\right.
\end{equation}
\begin{equation}\label{eq28}
	{\bar{f}}_{i,k}=\left\{\begin{aligned}
		&{{\kappa}_{1}} f_k^{\max }+\left(1-{{\kappa}_{1}}\right) {\bar{f}}_{i,k}, {{\kappa}_{2}}>0.5, \forall k \in \mathcal{K}, \\
		&{{\kappa}_{1}}{\vartheta} +\left(1-{{\kappa}_{1}}\right) {\bar{f}}_{i,k}, {{\kappa}_{2}} \leq 0.5, \forall k \in \mathcal{K},
	\end{aligned}\right.
\end{equation}
\begin{equation}\label{eq29}
	{\bar{p}}_{i,k}=\left\{\begin{aligned}
		&{{\kappa}_{1}} p_k^{\max }+\left(1-{{\kappa}_{1}}\right) {\bar{p}}_{i,k}, {{\kappa}_{2}}>0.5, \forall k \in \mathcal{K}, \\
		&{{\kappa}_{1}}{\vartheta} +\left(1-{{\kappa}_{1}}\right) {\bar{p}}_{i,k}, {{\kappa}_{2}} \leq 0.5, \forall k \in \mathcal{K},
	\end{aligned}\right.
\end{equation}
where ${{\kappa}_{1}}$ and ${{\kappa}_{2}}$ are random constants following 0-1 uniform distribution. The former is used for adjusting the mutation magnitudes of genes, but the latter is used for controlling their mutation directions. According to mutation rules \eqref{eq24}-\eqref{eq29}, we can easily find that the mutation operations of all genes will be performed in the feasible region.
\par
In order to ensure the diversity of population, and thus avoid the premature convergence of GA, a diversity-guided mutation is introduced before adaptive mutation and crossover operations. To this end, the diversity of the current population is measured. In the diversity-guided mutation, the mutation probability needs to be appropriately increased if the diversity level of the current population is low, it should be reduced otherwise. For n-dimensional numerical problems \cite{TZhou2022Oct}, the diversity measure is defined as
\begin{equation}\label{eq30}
	\begin{aligned}
		{\varsigma}=& \frac{1}{6}\left(\left(I{\mathscr{L}}_1\right)^{-1} \sum\nolimits_{i \in \mathcal{I}} \sqrt{\sum\nolimits_{k \in {\mathcal{K}}}\left(\bar{x}_{i,k}-\bar{x}_k^{\text{ave}}\right)^2}\right.\\
		&+\left(I{\mathscr{L}}_2\right)^{-1} \sum\nolimits_{i \in \mathcal{I}} \sqrt{\sum\nolimits_{k \in \mathcal{K}}\left(\bar{z}_{i,k}-\bar{z}_k^{\text{ave}}\right)^2} \\
          &+\left(I{\mathscr{L}}_3\right)^{-1} \sum\nolimits_{i \in \mathcal{I}} \sqrt{\sum\nolimits_{k \in \dot{\mathcal{K}}}\left(\bar{u}_{i,k}-\bar{u}_k^{\text{ave}}\right)^2} \\
		&+\left(I{\mathscr{L}}_4\right)^{-1} \sum\nolimits_{i \in \mathcal{I}} \sqrt{\sum\nolimits_{k \in \dot{\mathcal{K}}}({\bar{v}}_{i,k}-{\bar{v}}_k^{\text{ave}})^2} \\
		&+\left(I{\mathscr{L}}_5\right)^{-1} \sum\nolimits_{i \in \mathcal{I}} \sqrt{\sum\nolimits_{k \in \mathcal{K}}({\bar{f}}_{i,k}-{\bar{f}}_k^{\text{ave}})^2} \\
		&+\left.\left(I{\mathscr{L}}_6\right)^{-1} \sum\nolimits_{i \in \mathcal{I}} \sqrt{\sum\nolimits_{k \in \mathcal{K}}({\bar{p}}_{i,k}-{\bar{p}}_k^{\text{ave}})^2}\right),
	\end{aligned}
\end{equation}
\begin{equation}\label{eq31}
	\left\{\begin{aligned}
		 &\bar{x}_k^{\text{ave}}=\sum\nolimits_{i \in \mathcal{I}} \bar{x}_{i,k}/I, \forall k \in {\mathcal{K}}, \\
	 	&\bar{z}_k^{\text{ave}}=\sum\nolimits_{i \in \mathcal{I}} \bar{z}_{i,k}/I, \forall k \in \mathcal{K}, \\
           &\bar{u}_k^{\text{ave}}=\sum\nolimits_{i \in \mathcal{I}} \bar{u}_{i,k}/I, \forall k \in \dot{\mathcal{K}}, \\
		&{\bar{v}}_k^{\text{ave}}=\sum\nolimits_{i \in \mathcal{I}} {\bar{v}}_{i,k}/I, \forall k \in \dot{\mathcal{K}}, \\
		&{\bar{f}}_k^{\text{ave}}=\sum\nolimits_{i \in \mathcal{I}} {\bar{f}}_{i,k}/I, \forall k \in \mathcal{K}, \\
		&{\bar{p}}_k^{\text{ave}}=\sum\nolimits_{i \in \mathcal{I}} {\bar{p}}_{i,k}/I, \forall k \in \mathcal{K},
	\end{aligned}\right.
\end{equation}
where ${\mathscr{L}}_1$, ${\mathscr{L}}_2$, ${\mathscr{L}}_3$, ${\mathscr{L}}_4$, ${\mathscr{L}}_5$, and ${\mathscr{L}}_6$ are the diagonal lengths of feasible domains of $\mathbf{\bar{X}}$, $\mathbf{\bar{Z}}$, ${\mathbf{\bar{U}}}$, ${\mathbf{\bar{V}}}$, $\mathbf{\bar{F}}$, and $\mathbf{\bar{P}}$, respectively. Then, the diversity-guided mutation can be performed in the following probability.
\begin{equation}\label{eq32}
	\ddddot{\mathscr{P}}= \begin{cases}
        {\hbar}_7, & {\varsigma}<{\mu}_{1}, \\
        {\hbar}_8, & {\mu}_{1} \leq {\varsigma}<{\mu}_{2}, \\
        {\hbar}_9, & \text { other cases, }
        \end{cases}
\end{equation}
where $0<{\hbar}_7<1$, $0<{\hbar}_8<1$ and $0<{\hbar}_9<1$ are constant probabilities; ${\mu}_1$ and ${\mu}_{2}$ are the constant diversity thresholds.
\subsubsection{Elimination}
\par
To further enhance population diversity, some individuals with high similarity to others but low fitness will be eliminated and regenerated using initialization rules \eqref{eq21} at the latter stage of iteration. The similarity $\dot{\varsigma}_{i,i'}$ between any two individuals ${i}$ and $i'$ is defined as
\begin{equation}\label{eq32a}
	\dot{\varsigma}_{i,i'}=1/(1+\ddot{\varsigma}_{i,i'}),
\end{equation}
\begin{equation}\label{eq32b}
	\begin{aligned}
		&\ddot{\varsigma}_{i,i'}= \sqrt{\left({\sum\limits_{k \in {\mathcal{K}}}\left(\bar{x}_{i,k}-\bar{x}_{i',k}\right)^2}+{\sum\limits_{k \in \mathcal{K}}\left(\bar{z}_{i,k}-\bar{z}_{i',k}\right)^2}\right.} \\
          &\quad \quad +{\sum\nolimits_{k \in \dot{\mathcal{K}}}\left(\bar{u}_{i,k}-\bar{u}_{i',k}\right)^2} +{\sum\nolimits_{k \in \dot{\mathcal{K}}}({\bar{v}}_{i,k}-{\bar{v}}_{i',k})^2} \\
		&\quad \quad +\left.{\sum\nolimits_{k \in \mathcal{K}}({\bar{f}}_{i,k}-{\bar{f}}_{i',k})^2} + {\sum\nolimits_{k \in \mathcal{K}}({\bar{p}}_{i,k}-{\bar{p}}_{i',k})^2}\right),
	\end{aligned}
\end{equation}
where $\ddot{\varsigma}_{i,i'}$ is the Euclidean distance between individuals $i$ and $i'$.
\par
Up to now, the IADGGA can be summarized in Algorithm 1, where ${\mu}_{3}$ is similarity threshold.
\begin{table}[]
	\centering
	\begin{tabular}{ll}
		\toprule[1pt]
		\textbf{Algorithm 1: Improved Adaptive Diversity-Guided GA (IADGGA)} \\ \midrule[0.5pt]
        1: \textbf{Input:} $T_1$ (maximal number of iterations). \\
        2: \textbf{Output:} $\mathbf{\bar{X}}$, $\mathbf{\bar{U}}$, $\mathbf{\bar{Z}}$, $\mathbf{\bar{F}}$, ${\mathbf{\bar{V}}}$ and $\mathbf{\bar{P}}$ at $T_1$-th iteration. \\
		3: \textbf{Initialization:}\\
        4:\ \ \ \ Initialize iteration index: $t=1$.\\
		5:\ \ \ \ Establish the population of $I$ individuals using \eqref{eq21}.\\
		6:\ \ \ \ Evaluate the fitness of each individual using \eqref{eq20}.\\
		7:\ \ \ \ Identify the current best individual and replace the historical best\\
		8:\ \ \ \ \ \ with it if its fitness value surpasses the historical best.\\
		9: \textbf{While $t<=T_1$} \textbf{do}\\
		10:\ \ \ \ Generate a new population via the tournament method. \\
		11:\ \ \ \ \ \ Ensure the historically best individual is included; if not selected, \\
		12:\ \ \ \ \ \ replace the worst individual with it.\\
		13:\ \ \ Apply diversity-guided mutation using \eqref{eq24}-\eqref{eq29} under \eqref{eq32}.\\
		14:\ \ \ Re-evaluate the fitness of all individuals using \eqref{eq20}.\\
		15:\ \ \ Adaptively cross any two adjacent individuals with probability \eqref{eq22}.\\
        16:\ \ \ Apply adaptive mutation using \eqref{eq24}-\eqref{eq29} with probability \eqref{eq23}.\\
        17:\ \ \ \textbf{If} $t>=0.2T_1$ \textbf{do}\\
		18:\ \ \ \ Re-calculate the fitness values of all individuals  using \eqref{eq20}.\\
        19:\ \ \ \ Determine similarity between individuals ${i}$ and $i'$ using \eqref{eq32a}-\eqref{eq32b}.\\
        20:\ \ \ \ If $\dot{\varsigma}_{i,i'}>\mu$ for ${i}$ and $i'$, eliminate the one of them with a lower \\
        21:\ \ \ \ \ \ fitness value, and regenerated the eliminated one using \eqref{eq21}.\\
        22:\ \ \ \textbf{EndIf} \\
		23:\ \ \ Re-evaluate the fitness of all individuals using \eqref{eq20}.\\
		24:\ \ \ Identify the current best individual and replace the historical best\\
		25:\ \ \ \ \ with it if its fitness value surpasses the historical best.\\
		26:\ \ \ Update iteration index: ${t}={t}+1$.\\
		27: \textbf{EndWhile}\\ \bottomrule[0.5pt]
	\end{tabular}
	\label{alg1}
\end{table}
\subsection{APSO}
In PSO, any particle (individual) has two attributes including position and velocity, where the positions of particles represent the solutions to problem \eqref{eq19}, and their velocities show the evolutions of solutions. In this paper, after achieving the coarse-grained solutions to problem \eqref{eq19}, APSO is further used for finding its fine-grained ones. In APSO, the initial positions of particles are the outputs of IADGGA. According to the efforts in \cite{FVanddenbergh2007}, the velocity of any common particle (individual) ${i\in \mathcal{I}_{-i'}}$ can be updated by
\begin{equation}\label{eq33}
\ddot{x}_{i,k}^{t+1}= {\omega}_{i}^t \ddot{x}_{i,k}^t+{{\kappa}_{3}} \xi_{i,k}({\hat{x}}_{i,k}^t-{\dot{x}} _{i,k}^t)+{{\kappa}_{4}} \zeta_{i,k}({\tilde{x}} _k^t-{\dot{x}} _{i,k}^t), \forall k \in {\mathcal{K}},
\end{equation}
\begin{equation}\label{eq34}
\ddot{z}_{i,k}^{t+1}= {\omega}_{i}^t \ddot{z}_{i,k}^t+{{\kappa}_{3}} {\xi}_{i,k}({\hat{z}}_{i,k}^t-{\dot{z}} _{i,k}^t)+{{\kappa}_{4}} {\zeta}_{i,k}({\tilde{z}} _k^t-{\dot{z}} _{i,k}^t), \forall k \in \mathcal{K},
\end{equation}
\begin{equation}\label{eq35}
\ddot{u}_{i,k}^{t+1}= {\omega}_{i}^t \ddot{u}_{i,k}^t+{{\kappa}_{3}} \bar{\xi}_{i,k}({\hat{u}}_{i,k}^t-{\dot{u}} _{i,k}^t)+{{\kappa}_{4}} \bar{\zeta}_{i,k}({\tilde{u}} _k^t-{\dot{u}} _{i,k}^t), \forall k \in \dot{\mathcal{K}},
\end{equation}
\begin{equation}\label{eq36}
\ddot{v}_{i,k}^{t+1}= {\omega}_{i}^t \ddot{v}_{i,k}^t+{{\kappa}_{3}} \bar{\xi}_{i,k}({\hat{v}}_{i,k}^t-{\dot{v}} _{i,k}^t)+{{\kappa}_{4}} \bar{\zeta}_{i,k}({\tilde{v}} _k^t-{\dot{v}} _{i,k}^t), \forall k \in \dot{\mathcal{K}},
\end{equation}
\begin{equation}\label{eq37}
\ddot{f}_{i,k}^{t+1}= {\omega}_{i}^t \ddot{f}_{i,k}^t+{{\kappa}_{3}} {\xi}_{i,k}({\hat{f}}_{i,k}^t-{\dot{f}} _{i,k}^t)+{{\kappa}_{4}} {\zeta}_{i,k}({\tilde{f}} _k^t-{\dot{f}} _{i,k}^t), \forall k \in \mathcal{K},
\end{equation}
\begin{equation}\label{eq38}
\ddot{p}_{i,k}^{t+1}= {\omega}_{i}^t \ddot{p}_{i,k}^t+{{\kappa}_{3}} {\xi}_{i,k}({\hat{p}}_{i,k}^t-{\dot{p}} _{i,k}^t)+{{\kappa}_{4}} {\zeta}_{i,k}({\tilde{p}} _k^t-{\dot{p}} _{i,k}^t), \forall k \in \mathcal{K},
\end{equation}
where ${\omega}_{i}^t$ represents the inertia weight of particle ${i}$ at the $t$-th iteration; $0<{{\kappa}_{3}}<1$ and $0<{{\kappa}_{4}}<1$ represent self-learning and social learning factors, respectively; $\xi_{i,k}$, $\zeta_{i,k}$, $\bar{\xi}_{i,k}$ and $\bar{\zeta}_{i,k}$ are random numbers in the interval [0,1];  $\ddot{x}_{i,k}^{t}$, $\ddot{z}_{i,k}^{t}$, $\ddot{u}_{i,k}^{t}$, $\ddot{v}_{i,k}^{t}$, $ \ddot{f}_{i,k}^{t}$ and $\ddot{p}_{i,k}^{t}$ are the velocities of $\bar{x}_{i,k}^{t}$, $\bar{z}_{i,k}^{t}$, $\bar{u}_{i,k}^{t}$, $\bar{v}_{i,k}^{t}$,  $\bar{f}_{i,k}^{t}$ and $\bar{p}_{i,k}^{t}$ at $t$-th iteration, respectively; ${\dot{x}} _{i,k}^{t}$, ${\dot{z}} _{i,k}^{t}$, ${\bar{u}} _{i,k}^{t}$, ${\dot{v}} _{i,k}^{t}$, $ {\dot{f}} _{i,k}^{t}$ and ${\dot{p}} _{i,k}^{t}$ are the positions of $\bar{x}_{i,k}^{t}$, $\bar{z}_{i,k}^{t}$, $\bar{u}_{i,k}^{t}$, $\bar{v}_{i,k}^{t}$, $ \bar{f}_{i,k}^{t}$ and $\bar{p}_{i,k}^{t}$ at $t$-th iteration, respectively; $\mathbf{{\hat{X}}}_{i}^t=\{{\hat{x}}_{i,k}^t, \forall k \in \mathcal{K}\}$, $\mathbf{{\hat{Z}}}_{i}^t=\{{\hat{z}}_{i,k}^t, \forall k \in \mathcal{K}\}$, $\mathbf{{\hat{U}}}_{i}^t=\{{\hat{u}}_{i,k}^t, \forall k \in \dot{\mathcal{K}}\}$, $\mathbf{\hat{V}}_{i}^t=\{{\hat{v}}_{i,k}^t, \forall k \in \dot{\mathcal{K}}\}$, $\mathbf{{\hat{F}}}_{i}^t=\{{\hat{f}}_{i,k}^t, \forall k \in \mathcal{K}\}$ and $\mathbf{{\hat{P}}}_{i}^t=\{{\hat{p}}_{i,k}^t, \forall k \in \mathcal{K}\}$ represent the historically optimal position of particle ${i}$ at the $t$-th iteration; $\mathbf{{\tilde{x}} }^t=\{{\tilde{x}} _k^t, \forall k \in \dot{\mathcal{K}}\}$, $\mathbf{{\tilde{z}} }^{t}=\{{\tilde{z}} _k^t, \forall k \in \mathcal{M}\}$, $\mathbf{{\tilde{u}} }^t=\{{\tilde{u}} _k^t, \forall k \in \dot{\mathcal{K}}\}$, $\mathbf{\tilde{v}}_{i}^t=\{{\tilde{v}} _k^t, \forall k \in \dot{\mathcal{K}}\}$, $\mathbf{{\tilde{f}} }^t=\{\operatorname{{\tilde{f}} }_k^t, \forall k \in \mathcal{K}\}$ and $\mathbf{{\tilde{p}} }^t=\{{\tilde{p}} _k^t, \forall k \in \mathcal{K}\}$ represent the globally optimal position at the $t$-th iteration. Significantly, a particle at a historically optimal position is the personal best particle, the globally best particle refers to a particle whose fitness value is the highest among all personal best particles, and the position of a globally best particle is defined as the globally optimal position.
\par
In \eqref{eq33}-\eqref{eq38}, the inertia weight ${\omega}_{i}$ of any common particle ${i}$ can be updated by
\begin{equation}\label{eq39}
	{\omega}_{i}^{t+1}={\omega}_{i}^{t}-t\left({\omega}^{\max }-{\omega}^{\min }\right)/{T_2},
\end{equation}
where ${\omega}^{\max }$ and ${\omega}^{\min }$ are the minimum and maximum inertia weights respectively; $t$ and ${T_2}$ are the iteration index and number of iterations of APSO, respectively.
\par
After updating the velocities of particles, the position of any common particle ${i\in \mathcal{I}_{-i'}}$ can be updated by
\begin{equation}\label{eq40}
	{\dot{x}} _{i,k}^{t+1}=R({\dot{x}} _{i,k}^t+\ddot{x}_{i,k}^{t+1}), \forall k \in {\mathcal{K}},
\end{equation}
\begin{equation}\label{eq41}
	{\dot{z}} _{i,k}^{t+1}=R({\dot{z}} _{i,k}^t+\ddot{z}_{i,k}^{t+1}), \forall k \in \mathcal{K},
\end{equation}
\begin{equation}\label{eq42}
	{\dot{u}} _{i,k}^{t+1}=R({\dot{u}} _{i,k}^t+\ddot{u}_{i,k}^{t+1}), \forall k \in \dot{\mathcal{K}},
\end{equation}
\begin{equation}\label{eq43}
	{\dot{v}} _{i,k}^{t+1}=R({\dot{v}} _{i,k}^t+\ddot{v}_{i,k}^{t+1}), \forall k \in \dot{\mathcal{K}},
\end{equation}
\begin{equation}\label{eq44}
	{\dot{f}} _{i,k}^{t+1}={\dot{f}} _{i,k}^t+\ddot{f}_{i,k}^{t+1}, \forall k \in \mathcal{K},
\end{equation}
\begin{equation}\label{eq45}
	{\dot{p}} _{i,k}^{t+1}={\dot{p}} _{i,k}^t+\ddot{p}_{i,k}^{t+1}, \forall k \in \mathcal{K},
\end{equation}
\par
Then, the velocity of the globally optimal particle $i'$ can be updated by
\begin{equation}\label{eq46}
	\ddot{x}_{i',k}^{t+1}=-{\dot{x}} _{i',k}^t+{\tilde{x}} _k^t+{{\kappa}_{5}} \ddot{x}_{i',k}^t+{\dot{\omega}}^t(1-2 {a}_{i',k}), \forall k \in {\mathcal{K}},
\end{equation}
\begin{equation}\label{eq47}
	\ddot{z}_{i',k}^{t+1}=-{\dot{z}} _{i',k}^t+{\tilde{z}} _k^t+{{\kappa}_{5}} \ddot{z}_{i',k}^t+{\dot{\omega}}^t(1-2 {a}_{i',k}), \forall k \in {\mathcal{K}},
\end{equation}
\begin{equation}\label{eq48}
	\ddot{u}_{i',k}^{t+1}=-{\dot{u}} _{i',k}^t+{\tilde{u}} _k^t+{{\kappa}_{5}} \ddot{u}_{i',k}^t+{\dot{\omega}}^t(1-2 \bar{a}_{i',k}), \forall k \in \dot{\mathcal{K}},
\end{equation}
\begin{equation}\label{eq49}
	\ddot{v}_{i',k}^{t+1}=-{\dot{v}} _{i',k}^t+{\tilde{v}} _k^t+{{\kappa}_{5}} \ddot{v}_{i',k}^t+{\dot{\omega}}^t(1-2 \bar{a}_{i',k}), \forall k \in \dot{\mathcal{K}},
\end{equation}
\begin{equation}\label{eq50}
	\ddot{f}_{i',k}^{t+1}=-{\dot{f}} _{i',k}^t+{\tilde{f}} _k^t+{{\kappa}_{5}} \ddot{f}_{i',k}^t+{\dot{\omega}}^t(1-2 {a}_{i',k}), \forall k \in \mathcal{K},
\end{equation}
\begin{equation}\label{eq51}
	\ddot{p}_{i',k}^{t+1}=-{\dot{p}} _{i',k}^t+{\tilde{p}} _k^t+{{\kappa}_{5}} \ddot{p}_{i',k}^t+{\dot{\omega}}^t(1-2 {a}_{i',k}), \forall k \in \mathcal{K},
\end{equation}
where ${{\kappa}_{5}}$ is a constant coefficient; $0 \leq {a}_{i',k} \leq 1$ and $0 \leq \bar{a}_{i',k} \leq 1$ are random numbers following 0-1 uniform distribution; ${\dot{\omega}}^t$ is the scaling factor at the $t$-th iteration;
\par
Then, the position of the globally optimal particle $i'$ is updated by
\begin{equation}\label{eq52}
	{\dot{x}} _{i',k}^{t+1}=R({\tilde{x}} _k^t+{{\kappa}_{5}} \ddot{x}_{i',k}^t+{\dot{\omega}}^t(1-2 {a}_{i',k})), \forall k \in {\mathcal{K}},
\end{equation}
\begin{equation}\label{eq53}
	{\dot{z}} _{i',k}^{t+1}=R({\tilde{z}} _k^t+{{\kappa}_{5}} \ddot{z}_{i',k}^t+{\dot{\omega}}^t(1-2 {a}_{i',k})), \forall k \in \mathcal{K},
\end{equation}
\begin{equation}\label{eq54}
	{\dot{u}} _{i',k}^{t+1}=R({\tilde{u}} _k^t+{{\kappa}_{5}} \ddot{u}_{i',k}^t+{\dot{\omega}}^t(1-2 \bar{a}_{i',k})), \forall k \in \dot{\mathcal{K}},
\end{equation}
\begin{equation}\label{eq55}
	{\dot{v}} _{i',k}^{t+1}=R({\tilde{v}} _k^t+{{\kappa}_{5}} \ddot{v}_{i',k}^t+{\dot{\omega}}^t(1-2 \bar{a}_{i',k})), \forall k \in \dot{\mathcal{K}},
\end{equation}
\begin{equation}\label{eq56}
	{\dot{f}} _{i',k}^{t+1}={\tilde{f}} _k^t+{{\kappa}_{5}} \ddot{f}_{i',k}^t+{\dot{\omega}}^t(1-2 {a}_{i',k}), \forall k \in \mathcal{K},
\end{equation}
\begin{equation}\label{eq57}
	{\dot{p}} _{i',k}^{t+1}={\tilde{p}} _k^t+{{\kappa}_{5}} \ddot{p}_{i',k}^t+{\dot{\omega}}^t(1-2 {a}_{i',k}), \forall k \in \mathcal{K}.
\end{equation}
In \eqref{eq52}-\eqref{eq57}, ${\dot{\omega}}^{t+1}$ is updated by
\begin{equation}\label{eq58}
	{\dot{\omega}}^{t+1}= \begin{cases}
2 {\dot{\omega}}^t, & \dot{n}>{{\mu}_{4}}, \\
0.5 {\dot{\omega}}^t, & \ddot{n}>{\mu}_{5}, \\
{\dot{\omega}}^t, & \text { other cases,}
\end{cases}
\end{equation}
where $\dot{n}$ represents the number of consecutive successes, but $\ddot{n}$ is the one of successive failures; ${{\mu}_{4}}$ and ${\mu}_{5}$ are the thresholds. Significantly, $G(\mathbf{\tilde{\Xi}}^t)=G(\mathbf{\tilde{\Xi}}^{t-1})$ is defined as failure and other cases are defined as success, where $\mathbf{\tilde{\Xi}}=\{\mathbf{{\tilde{x}} },\mathbf{{\tilde{z}} },\mathbf{{\tilde{u}} },\mathbf{{\tilde{v}} },\mathbf{{\tilde{f}} },\mathbf{{\tilde{p}} }\}$.
\begin{table}[]
	\centering
	\begin{tabular}{ll}
		\toprule[1pt]
		\textbf{Algorithm 2: Adaptive PSO (APSO)} \\ \midrule[0.5pt]
        1: \textbf{Input:} Parameter $T_2$, $\mathbf{\bar{X}}$, $\mathbf{\bar{Z}}$, $\mathbf{\bar{U}}$, ${\mathbf{\bar{V}}}$, $\mathbf{\bar{F}}$ and $\mathbf{\bar{P}}$ at $T_1$-th iteration. \\
        2: \textbf{Output:} Globally optimal position, i.e., $\mathbf{\tilde{\Xi}}$ at $T_2$-th iteration.  \\
		3: \textbf{Initialization:}\\
        4:\ \ \ \ Initialize iteration index: $t=1$.\\
        5:\ \ \ \ Initialize ${\ddot{\mathbf{X}}^{t}}$, ${\ddot{\mathbf{Z}}^{t}}$, ${\ddot{\mathbf{U}}^{t}}$, ${\ddot{\mathbf{V}}^{t}}$, ${\ddot{\mathbf{F}}^{t}}$ and ${\ddot{\mathbf{P}}^{t}}$ using random numbers\\
        6:\ \ \ \ \ \ following 0-1 uniform distribution.\\
        7:\ \ \ \ Set ${\dot{{\mathbf{X}} }^{t}}={\mathbf{{\hat{X}}}^{t}}={\mathbf{\bar{X}}}$, ${\dot{{\mathbf{Z}} }^{t}}={\mathbf{{\hat{Z}}}^{t}}={\mathbf{\bar{Z}}}$, ${\dot{{\mathbf{U}} }^{t}}={\mathbf{{\hat{U}}}^{t}}={\mathbf{\bar{U}}}$,\\
        8:\ \ \ \ \ \ ${\dot{{\mathbf{V}} }^{t}}={\mathbf{{\hat{V}}}^{t}}={\mathbf{\bar{V}}}$, ${\dot{{\mathbf{F}} }^{t}}={\mathbf{{\hat{F}}}^{t}}={\mathbf{\bar{F}}}$, ${\dot{{\mathbf{P}} }^{t}}={\mathbf{{\hat{P}}}^{t}}={\mathbf{\bar{P}}}$.\\
		9:\ \ \ \ Find the globally best particle.\\
		10: \textbf{While $t<=T_2$} \textbf{do}\\
		11:\ \ \ \ Update inertia weight using \eqref{eq39}.\\
		12:\ \ \ \ Update velocities of all common particles via \eqref{eq33}-\eqref{eq38}.\\
		13:\ \ \ \ Update positions of all common particles via \eqref{eq40}-\eqref{eq45}.\\
		14:\ \ \ \ Update velocity of globally best particle via \eqref{eq46}-\eqref{eq51}.\\
		15:\ \ \ \ Update position of globally best particle via \eqref{eq52}-\eqref{eq57}.\\
		16:\ \ \ \ Evaluate the fitness of each particle via \eqref{eq20}.\\
		17:\ \ \ \ For each particle, update the personal best attributes with the \\
		18:\ \ \ \ \ \ current attributes if the current fitness value surpasses the \\
		19:\ \ \ \ \ \ historically optimal value.\\
		20:\ \ \ \ Identify the globally best particle.\\
		21:\ \ \ \ Update scaling factor ${\dot{\omega}}$ via \eqref{eq58}.\\
		22:\ \ \ \ Update iteration index: $t=t+1$.\\
		23: \textbf{EndWhile}\\ \bottomrule[0.5pt]
	\end{tabular}
	\label{alg2}
\end{table}
\par
Up to now, the whole procedure of APSO can be summarized in Algorithm 2. In such an algorithm, $\dot{{\mathbf{X}} }^{t}=\{ {{\dot{x}} _{i,k}^{t}}, \forall i\in \mathcal{I},\forall k\in \mathcal{{K}} \}$, $\dot{{\mathbf{Z}} }^{t}=\{ {{\dot{z}} _{i,k}^{t}},\forall i\in \mathcal{I},\forall k\in \mathcal{K} \}$, $\dot{{\mathbf{U}} }^{t}=\{ {{\dot{u}} _{i,k}^{t}},\forall i\in \mathcal{I},\forall k\in \dot{\mathcal{K}} \}$, $\dot{{\mathbf{V}} }^{t}=\{ {\dot{v}} _{i,k}^{t},\forall i\in \mathcal{I},\forall k\in \dot{\mathcal{K}} \}$, $\dot{{\mathbf{F}} }^{t}=\{ {{\dot{f}} _{i,k}^{t}},\forall i\in \mathcal{I},\forall k\in \mathcal{K} \}$, $\dot{{\mathbf{P}} }^{t}=\{ {{\dot{p}} _{i,k}^{t}},\forall i\in \mathcal{I},\forall k\in \mathcal{K} \}$; $\ddot{\mathbf{X}}^{t}=\{ {\ddot{x}_{i,k}^{t}}, \forall i\in \mathcal{I},\forall k\in \mathcal{{K}} \}$, $\ddot{\mathbf{Z}}^{t}=\{ {\ddot{z}_{i,k}^{t}},\forall i\in \mathcal{I},\forall k\in \mathcal{K} \}$, $\ddot{\mathbf{U}}^{t}=\{ {\ddot{u}_{i,k}^{t}},\forall i\in \mathcal{I},\forall k\in \dot{\mathcal{K}} \}$, $\ddot{\mathbf{V}}^{t}=\{ \ddot{v}_{i,k}^{t},\forall i\in \mathcal{I},\forall k\in \dot{\mathcal{K}} \}$, $\ddot{\mathbf{F}}^{t}=\{ {\ddot{f}_{i,k}^{t}},\forall i\in \mathcal{I},\forall k\in \mathcal{K} \}$, $\ddot{\mathbf{P}}=\{ {\ddot{p}_{i,k}},\forall i\in \mathcal{I},\forall k\in \mathcal{K} \}$.
\par
Finally, the entire methodology devised to address problem \eqref{eq19} can be condensed into Algorithm 3. It is evident that this is a single-layer iterative algorithm. Initially, it conducts a coarse-grained search utilizing Algorithm 1, followed by a fine-grained search using Algorithm 2.
\begin{table}[]
	\centering
	\begin{tabular}{ll}
		\toprule[1pt]
		\textbf{Algorithm 3: Further Improved HAS (FIHAS)} \\ \midrule[0.5pt]
    1: \textbf{Input:} Parameters of ADGGA and APSO. \\
    2: \textbf{Output:} Globally optimal position, i.e., $\mathbf{\tilde{\Xi}}$ at $T_2$-th iteration. \\
	3: Determine coarse-grained solutions by leveraging Algorithm 1.\\
	4: Obtain fine-grained solutions by employing Algorithm 2.\\
	 \bottomrule[0.5pt]
	\end{tabular}
	\label{alg3}
\end{table}
\section{Algorithm Analysis}\label{sec5}
\subsection{Convergence analysis}
\par
\noindent
\textbf{\textit{Theorem 1:}} ADGGA is globally convergent.
\par
\textit{Proof:}  In ADGGA, a tournament-based approach is employed to choose $I$ individuals, following which the worst-performing individual among the selected is consistently swapped with the historically best one. This ensures that the historically best individual remains present after the selection process. According to \cite{MYLi2004Sep}, GA can achieve global optimum convergence when they consistently retain the best individual both before and after selection. Therefore, it follows that ADGGA is capable of converging to the global optimum. \ding{113}
\par
When $\ddddot{\mathscr{P}}=0$, ADGGA becomes traditional adaptive GA. Then, some results related to Theorem 1 can be easily established as follows.
\par
\noindent
\textit{\textbf{Corollary 1:}} ADGGA, with the parameter $\ddddot{\mathscr{P}}=0$ set to 0, converges to the global optimum by ensuring that the best individual is always preserved both before and after the selection process.
\par
Next, some investigations on the convergence of APSO will be done. To this end, some important results need to be first provided.
\par
\noindent
\textit{\textbf{Lemma 1:}} APSO meets the condition H1 in \cite{FVanddenbergh2007}.
\par
\textit{Proof}: At first, a function ${{\mathcal{G}}_{1}}$ is defined as
\begin{equation}\label{eq59}
	{{\mathcal{G}}_{1}}( {{\mathbf{\tilde{\Xi}}}^{t}},\dot{\mathbf{\Xi}}_{i}^{t} )=\left\{ \begin{split}
		& {{\mathbf{\tilde{\Xi}}}^{t}},\ \text{if}\ {G}(\dot{\mathbf{\Xi}}_{i}^{t+1})\le {G}(  {{\mathbf{\tilde{\Xi}}}^{t}} ), \\
		& \dot{\mathbf{\Xi}}_{i}^{t+1},\ \text{otherwise}, \\
	\end{split} \right.	
\end{equation}	
where $\dot{\mathbf{\Xi}}_{i}^{t+1}=\{\dot{{\mathbf{X}} }_{i}^{t+1},\dot{{\mathbf{Z}} }_{i}^{t+1},\dot{{\mathbf{U}} }_{i}^{t+1},\dot{{\mathbf{V}} }_{i}^{t+1},\dot{{\mathbf{F}} }_{i}^{t+1},\dot{{\mathbf{P}} }_{i}^{t+1} \}$;
\begin{equation}\label{eq60}
\dot{{\mathbf{X}} }_{i}^{t+1}=R(\dot{{\mathbf{X}}}_{i}^{t}+{\omega}_{i}^t\ddot{\mathbf{X}}_{i}^{t}+{{\kappa}_{3}}{\boldsymbol{\xi}_{i}}( \mathbf{{\hat{X}}}_{i}^{t}-\dot{{\mathbf{X}} }_{i}^{t})+{{\kappa}_{4}}{{\boldsymbol{\zeta}}_{i}}(\mathbf{{\tilde{x}} }^{t}-\dot{{\mathbf{X}} }_{i}^{t})),
\end{equation}
\begin{equation}\label{eq61}
\dot{{\mathbf{Z}} }_{i}^{t+1}=R(\dot{{\mathbf{Z}} }_{i}^{t}+{\omega}_{i}^t\ddot{\mathbf{Z}}_{i}^{t}+{{\kappa}_{3}}{{\boldsymbol{\xi}}_{i}}( \mathbf{{\hat{Z}}}_{i}^{t}-\dot{{\mathbf{Z}} }_{i}^{t})+{{\kappa}_{4}}{{{\boldsymbol{\zeta}}}_{i}}(\mathbf{{\tilde{z}} }^{t}-\dot{{\mathbf{Z}} }_{i}^{t})),
\end{equation}
\begin{equation}\label{eq62}
\dot{{\mathbf{U}} }_{i}^{t+1}=R(\dot{{\mathbf{U}} }_{i}^{t}+{\omega}_{i}^t\ddot{\mathbf{U}}_{i}^{t}+{{\kappa}_{3}}{\bar{\boldsymbol{\xi}}_{i}}( \mathbf{{\hat{U}}}_{i}^{t}-\dot{{\mathbf{U}} }_{i}^{t})+{{\kappa}_{4}}{{\bar{\boldsymbol{\zeta}}}_{i}}(\mathbf{{\tilde{u}} }^{t}-\dot{{\mathbf{U}} }_{i}^{t})),
\end{equation}
\begin{equation}\label{eq63}
\dot{{\mathbf{V}} }_{i}^{t+1}=R(\dot{{\mathbf{V}} }_{i}^{t}+{\omega}_{i}^t\ddot{\mathbf{V}}_{i}^{t}+{{\kappa}_{3}}{\bar{\boldsymbol{\xi}}_{i}}( \mathbf{{\hat{V}}}_{i}^{t}-\dot{{\mathbf{V}} }_{i}^{t})+{{\kappa}_{4}}{{\bar{\boldsymbol{\zeta}}}_{i}}(\mathbf{{\tilde{v}} }^{t}-\dot{{\mathbf{V}} }_{i}^{t})),
\end{equation}
\begin{equation}\label{eq64}
	\dot{{\mathbf{F}} }_{i}^{t+1}=\dot{{\mathbf{F}} }_{i}^{t}+{\omega}_{i}^t\ddot{\mathbf{F}}_{i}^{t}+{{\kappa}_{3}}{{\boldsymbol{\xi}}_{i}}( \mathbf{{\hat{F}}}_{i}^{t}-\dot{{\mathbf{F}} }_{i}^{t} )+{{\kappa}_{4}}{{{\boldsymbol{\zeta}}}_{i}}(\mathbf{{\tilde{f}} }^{t}-\dot{{\mathbf{F}} }_{i}^{t}),
\end{equation}
\begin{equation}\label{eq65}
\dot{{\mathbf{P}} }_{i}^{t+1}=\dot{{\mathbf{P}} }_{i}^{t}+{\omega}_{i}^t\ddot{\mathbf{P}}_{i}^{t}+{{\kappa}_{3}}{{\boldsymbol{\xi}}_{i}}( \mathbf{{\hat{P}}}_{i}^{t}-\dot{{\mathbf{P}} }_{i}^{t} )+{{\kappa}_{4}}{{{\boldsymbol{\zeta}}}_{i}}(\mathbf{{\tilde{p}} }^{t}-\dot{{\mathbf{P}} }_{i}^{t}).
\end{equation}
In \eqref{eq60}-\eqref{eq65}, ${\boldsymbol{\xi }_{i}}=\{ {{\xi}_{i,k}},\forall k\in {\mathcal{K}} \}$, ${\boldsymbol{\zeta}_{i}}=\{ {{\zeta}_{i,k}}, k\in {\mathcal{K}} \}$, ${\bar{\boldsymbol{\xi }}_{i}}=\{ {\bar{\xi}_{i,k}},\forall k\in \dot{\mathcal{K}} \}$ and ${\bar{\boldsymbol{\zeta}}_{i}}=\{ {\bar{\zeta}_{i,k}}, k\in \dot{\mathcal{K}} \}$. Then, we can easily deduce ${G}( {{\mathcal{G}}_{1}}( \mathbf{\tilde{\Xi}}^{t},\dot{\mathbf{\Xi}}_{i}^{t} ) )\ge {G}( \mathbf{\tilde{\Xi}}^{t} )$. Evidently, the common PSO consisting of \eqref{eq33}-\eqref{eq45} meets the condition H1 in \cite{FVanddenbergh2007}.
\par
Next, we concentrate on the update \eqref{eq46}-\eqref{eq57}. At first, another function $\mathcal{G}_{2}$ is defined as
\begin{equation}\label{eq66}
	{{\mathcal{G}}_{2}}( \mathbf{\tilde{\Xi}}^{t},\dot{\mathbf{\Xi}}_{i'}^{t} )=\left\{ \begin{split}
		& \mathbf{\tilde{\Xi}}^{t},\ \text{if}\ {G}( \dot{\mathbf{\Xi}}_{i'}^{t+1} )\le {G}(\mathbf{\tilde{\Xi}}^{t} ), \\
		& \dot{\mathbf{\Xi}}_{i'}^{t+1}, \text{otherwise}, \\
	\end{split} \right.
\end{equation}
where  $\dot{\mathbf{\Xi}}_{i'}^{t+1}=\{\dot{{\mathbf{X}} }_{i'}^{t+1},\dot{{\mathbf{U}} }_{i'}^{t+1},\dot{{\mathbf{Z}} }_{i'}^{t+1},\dot{{\mathbf{F}} }_{i'}^{t+1},\dot{{\mathbf{V}} }_{i'}^{t+1},\dot{{\mathbf{P}} }_{i'}^{t+1} \}$ ;
\begin{equation}\label{eq67}
	\dot{{\mathbf{X}} }_{i'}^{t+1}=R( \mathbf{{\tilde{x}} }^{t}+{{\kappa}_{5}}\ddot{\mathbf{X}}_{i'}^{t}+{{\dot{\omega}}^{t}}( 1-2{\mathbf{A}_{i'}} ) ),
\end{equation}
\begin{equation}\label{eq68}
	\dot{{\mathbf{Z}} }_{i'}^{t+1}=R( \mathbf{{\tilde{z}} }^{t}+{{\kappa}_{5}}\ddot{\mathbf{Z}}_{i'}^{t}+{{\dot{\omega}}^{t}}( 1-2{\mathbf{A}_{i'}} ) ),
\end{equation}
\begin{equation}\label{eq69}
	\dot{{\mathbf{U}} }_{i'}^{t+1}=R( \mathbf{{\tilde{u}} }^{t}+{{\kappa}_{5}}\ddot{\mathbf{U}}_{i'}^{t}+{{\dot{\omega}}^{t}}( 1-2{\mathbf{\bar{A}}_{i'}} ) ),
\end{equation}
\begin{equation}\label{eq70}
	\dot{{\mathbf{V}} }_{i'}^{t+1}=R( \mathbf{{\tilde{v}} }^{t}+{{\kappa}_{5}}\ddot{\mathbf{V}}_{i'}^{t}+{{\dot{\omega}}^{t}}( 1-2{\mathbf{\bar{A}}_{i'}} )),
\end{equation}
\begin{equation}\label{eq71}
	\dot{{\mathbf{F}} }_{i'}^{t+1}=\mathbf{{\tilde{f}} }^{t}+{{\kappa}_{5}}\ddot{\mathbf{F}}_{i'}^{t}+{{\dot{\omega}}^{t}}( 1-2{{\mathbf{A}}_{i'}} ),
\end{equation}
\begin{equation}\label{eq72}
	\dot{{\mathbf{P}} }_{i'}^{t+1}=\mathbf{{\tilde{p}} }^{t}+{{\kappa}_{5}}\ddot{\mathbf{P}}_{i'}^{t}+{{\dot{\omega}}^{t}}( 1-2{{\mathbf{A}}_{i'}} ),
\end{equation}
In \eqref{eq67}-\eqref{eq72},  ${\mathbf{A}_{i'}}=\{ {{a}_{i',k}},\forall k\in {\mathcal{K}} \}$ and ${\mathbf{\bar{A}}_{i'}}=\{ {\bar{a}_{i',k}},\forall k\in \dot{\mathcal{K}} \}$.
\par
It is easy to deduce ${G}( {{\mathcal{G}}_{2}}( \mathbf{\tilde{\Xi}}^{t},\dot{\mathbf{\Xi}}_{i'}^{t} ) )\ge {G}( \mathbf{\tilde{\Xi}}^{t} )$. From the above-mentioned analyses, it is easy to infer that APSO finally meets condition H1 in \cite{FVanddenbergh2007}. \ding{113}
\par
\noindent
\textbf{\textit{Lemma 2:}} APSO satisfies condition H3 in  \cite{FVanddenbergh2007}.
\par
\textit{Proof:} From the results in \cite{FVanddenbergh2007}, it is easy to infer that APSO meets condition H3 in  \cite{FVanddenbergh2007}. To this end, $\mathcal{D}_{0}$ and $ \mathcal{D}_{\delta}$ are defined at first. For any $ \mathbf{\hat{\Xi}}_{0}\in \mathcal{R} $, $ \mathcal{D}_{0}=\{\mathbf{\hat{\Xi}}\in\mathcal{R} | {G}( \mathbf{\hat{\Xi}} ) \ge {G}( {{\mathbf{\hat{\Xi}}}_{0}} ) \}$, where $ \mathcal{R} $ denotes the feasible region of $\dot{\mathbf{\Xi}}$. In addition, $ \mathcal{D}_{\delta}=\{\mathbf{\hat{\Xi}}\in\mathcal{R} | {G}( \mathbf{\hat{\Xi}} ) \ge \delta +\chi \} $, where $\delta\textgreater{0}$, $\chi=\sup( \mathcal{A}:{\phi}[ \mathbf{\hat{\Xi}}\in \mathcal{R}| {G}( \mathbf{\hat{\Xi}} )>\mathcal{A} ]>0) $ is an essential supremum of ${G}(\mathbf{\hat{\Xi}})$ under Lebesgue measure ${\phi}[ \mathbf{\hat{\Xi}}\in \mathcal{R}| {G}( \mathbf{\hat{\Xi}} )>\mathcal{A} ]>0$.
\par
We assume that $\mathbf{\tilde{\Xi}}$ belongs to $ \mathcal{D}_{0} $. According to the principles of APSO, it is straightforward to deduce that the sampling consistently occurs in the vicinity of a point within $ \mathcal{D}_{0} $. Furthermore, when initiating from any point in $\mathcal{D}_{0}$, APSO consistently guarantees a non-degenerate sampling volume with a non-zero probability. This probability pertains to the likelihood of sampling a point that is nearer to the optimal region $\mathcal{D}_{\delta}$. Based on these analyses, it is simple to conclude that APSO satisfies condition H3 as outlined in \cite{FVanddenbergh2007}. \ding{113}
\par
\noindent
\textbf{\textit{Theorem 2:}} APSO converges to the local minimum.
\par
\textit{Proof:} As indicated in \cite{FVanddenbergh2007}, the convergence of APSO to a local minimum is attributed to its fulfillment of conditions H1 and H3. \ding{113}

\subsection{Complexity Analysis}
The computation complexity of Algorithms 1, 2 and 3 will be analyzed as follows.
\par
\noindent
\textbf{\textit{Proposition 1}}: The computation complexity of Algorithm 1 is $\mathcal{O}(\max({MKI^2}T_1,{SIK^2}T_1))$.
\par
\textit{Proof:} In Steps 5, 20-21, the computation complexity mainly comes from the initialization of the longest genes including $\mathbf{\bar{X}}$, $\mathbf{\bar{U}}$, ${\mathbf{\bar{V}}}$, and it is $\mathcal{O}( {MIK} )$. In Steps 7-8 and 24-25, the computation complexity is $\mathcal{O}( {I} )$. In Steps 10-12, the computation complexity mainly comes from the selection operation. According to the analyses in \cite{FGuo2018Dec}, it has a complexity of $\mathcal{O}( {2I} )$ since two individuals are selected in each round of tournament method. In Steps 13 and 16, the computation complexity mainly comes from the mutation of the longest genes, and it is $\mathcal{O}({MIK} )$ for all individuals. In Step 15, the computation complexity is $\mathcal{O}( {I} )$ for all individuals.
\par
In Steps 6, 14, 18 and 23, uplink data rates can be calculated before executing these steps. The computation complexity of these rates is highly dependent on the calculation of uplink interferences. To reduce the computation complexity, we just need to calculate interferences received from other IMDs on the same utilized channels, and do it before calculating uplink rates. Evidently, the number of associated IMDs on a channel is often less than or equal to $K$, and the highest computation complexity used for calculating interferences is $\mathcal{O}( {SK^2} )$. Consequently, the highest computation complexity of uplink data rates is $\mathcal{O}( {SK^2} )$. Significantly, we utilize codes for the calculation of fitness values. That is to say, the indices of channel and BS selected by virtual IMDs are utilized, and indices of cryptographic algorithms selected by tasks are also used. Based on these, the computation complexity of other operations done for the calculation of fitness values is $\mathcal{O}( {MK} )$. In general, the computation complexity of Steps 6, 14, 18 and 23 is the maximum between $\mathcal{O}({MK} )$ and $\mathcal{O}({SK^2} )$ for each individual, and is the maximum between $\mathcal{O}({MIK} )$ and $\mathcal{O}({SIK^2} )$ for all individuals. In Step 19, the computation complexity mainly comes from the calculation for the longest genes, and it is $\mathcal{O}({MKI^2} )$ for all individuals.
\par
In general, the computation complexity of Algorithm 1 is the maximum between $\mathcal{O}({MKI^2}T_1)$ and $\mathcal{O}({SIK^2}T_2)$. \ding{113}
\par
\noindent
\textbf{\textit{Proposition 2}}: The computation complexity of Algorithm 2 is $\mathcal{O}(\max({MIK}T_2,{MSK^2}T_2))$.
\par
\emph{Proof}: In Steps 4-8 of Algorithm 2, the computation complexity mainly derives from the initialization of the longest genes including $\dot{{\mathbf{X}} }$, $\dot{{\mathbf{U}} }$, $\dot{{\mathbf{V}} }$, which is $\mathcal{O}(MIK )$. In Step 9, the computation complexity is $\mathcal{O}(I)$. In Steps 12-15, the computation complexity mainly derives from the update of positions and velocities of the longest genes, which is $\mathcal{O}(MIK )$. As revealed in Algorithm 1, the computation complexity of Step 16 is $\mathcal{O}( \max({MIK}, {SIK^2}))$ for all individuals. In addition, it is evident that the computation complexity of Steps 17-22 is $\mathcal{O}(I)$.
\par
In general, the computation complexity of Algorithm 2 is the maximum between $\mathcal{O}({MIK}T_2)$ and $\mathcal{O}({SIK^2}T_2)$ after $T_2$ iterations. \ding{113}
\par
\noindent
\textbf{\textit{Proposition 3}}: The computation complexity of Algorithm 3 is $\mathcal{O}(\max({MKI^2}T_1,{SIK^2}T_1,{MIK}T_2,{SIK^2}T_2))$.
\par
\textit{Proof:} In Algorithm 3, Algorithms 1 and 2 are executed sequentially. Therefore, the computation complexity of Algorithm 3 is the maximum among $\mathcal{O}({MKI^2}T_1)$, $\mathcal{O}({SIK^2}T_1)$, $\mathcal{O}({MIK}T_2)$ and $\mathcal{O}({SIK^2}T_2)$. \ding{113}

\section{NUMERICAL RESULTS}\label{sec6}
In cache-assisted ultra-dense MEC networks with multi-slope channels, the noise power spectrum density is -174 dBm/Hz, $W=20$ MHz, $N=30$, $K=20$, $L=6$, $M=3$, $f_k^{\text{lmax}}=1\sim2$ GHz, $p_k^{\text{max}}=23$ dBm, and $f_n^{\text{mmax}}=20$ GHz \cite{TZhou2022Oct}; $\ddot{h}_{j}=0.3$ km, $\gamma_j^\text{LS}=2.09$, $\gamma_j^{\mathrm{\text{NLS}}}=3.75$, ${h}_j^{\mathrm{L}}=10^{-10.38}$, and ${h}_j^{\mathrm{\text{NLS}}}=10^{-14.54}$ \cite{WTeng2021Spet}; $\alpha=10^{-24}$, $\tau_{k,m}^{\text{max}}=1\sim3$ s, ${\theta}_{k,m}=1\sim3$, $\lambda_{k,m}=5\times10^{3}\sim10\times10^{3}$ USD, $d_{k,m}=0.01\sim0.05$ MB, and ${\ell}_{k,m}=10\sim50$ Mcycles; $\rho_{k,m}=3\sim6$; $\boldsymbol{\epsilon}=[100, 200, 250, 300, 350, 1050]$ cycles/bit, $\boldsymbol{\dot{\epsilon}}=[90, 280, 350, 300, 400, 1700]$ cycles/bit, and $\boldsymbol{\ddot{\epsilon}}=[2.5296, 5.0425, 6.837, 7.8528, 8.7073, 26.3643]\times10^{-7}$ J/bit \cite{YZhang2020Nov}; ${\hbar}_{1}=0.8$, ${\hbar}_{2}=0.3$, ${\hbar}_{3}=0.1$, ${\hbar}_{4}=2.5$, ${\hbar}_{5}=0.01$, ${\hbar}_{6}=0.993$, and ${\hbar}_{7}=0.6$ \cite{YZhou2022}; ${\hbar}_{8}=0.03$, ${\hbar}_{9}=1\times10^{-5}$, ${\mu}_{1}=0.01$, ${\mu}_{2}=0.25$, ${{{\mu}_{4}}}=15$, ${{\mu}_{5}}=5$, ${{\kappa}_{3}}={{\kappa}_{4}}={{\kappa}_{5}}=2$, ${\omega}^{\max }=0.9$, ${\omega}^{\min }=0.4$ \cite{TZhou2022Oct}.
\par
To underscore the efficiency of FIHAS, the subsequent algorithms are introduced as comparative benchmarks.
\par
\noindent
\textbf{\textit{Completely Local Computing Algorithm (CLCA):}} To reduce the energy consumed by IMDs, all computing tasks of IMDs are executed by themselves with computing capacity $\min({{\ell}_{k,m}}/\tau_{k,m}^{\text{max}},{f}_{k}^{\text{lmax}})$ locally. That is to say, all IMDs may complete computing tasks within the maximum tolerable delays with computing capacity less than or equal to their maximum allowable capacity.
\par
\noindent
\textbf{\textit{Completely Offloading Algorithm (COA):}} All computational tasks of IMDs are transferred to BSs with the highest channel gains for execution. In COA, cryptographic algorithms with the lowest security breach costs are chosen for tasks, and the computational resources of the BSs are allocated to the tasks in proportion to their respective CPU occupation ratios.
\par
\noindent
\textbf{\textit{Improved Hierarchical Adaptive Search (IHAS) \cite{TZhou2022Oct}:}} To solve the problem \eqref{eq19}, IHAS is introduced.
\begin{figure}[!t]
	\centerline{\includegraphics[width=3.8in]{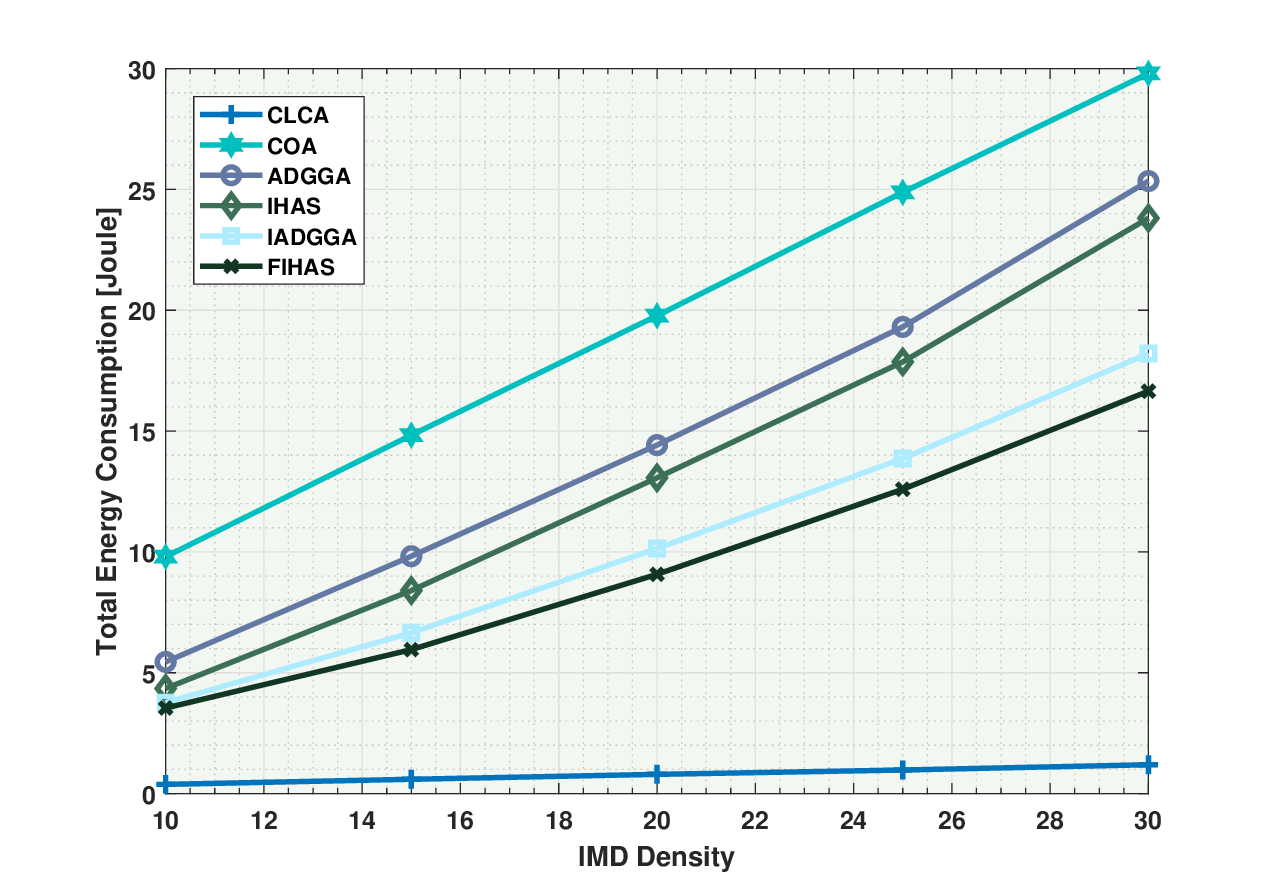}}
	\caption{Influences of IMD density on the total energy consumption (TEC).}
	\label{fig4}
\end{figure}
\par
Fig.\ref{fig4} shows the influences of IMD density on the total energy consumption (TEC), where the IMD density refers to the number of IMDs at each macrocell. As illustrated in Fig.\ref{fig4}, TEC should increase with IMD density since more IMDs often consume more energy. Among all offloading algorithms, COA may achieve the highest TEC since the best-channel association often results in extremely unbalanced load distribution and high computation time, but CLCA may achieve the lowest TEC since it has no energy consumption used for transmission, encryption and decryption. In addition, IADGGA may achieve a lower TEC than ADGGA since the former adopts more effective crossover and mutation probabilities to more fully search solution space, and maintains lowly-similar individuals while mutating highly-similar individuals to enhance population diversity. Similarly, FIHAS may achieve a lower TEC than IHAS.
\begin{figure}[!t]
	\centerline{\includegraphics[width=3.8in]{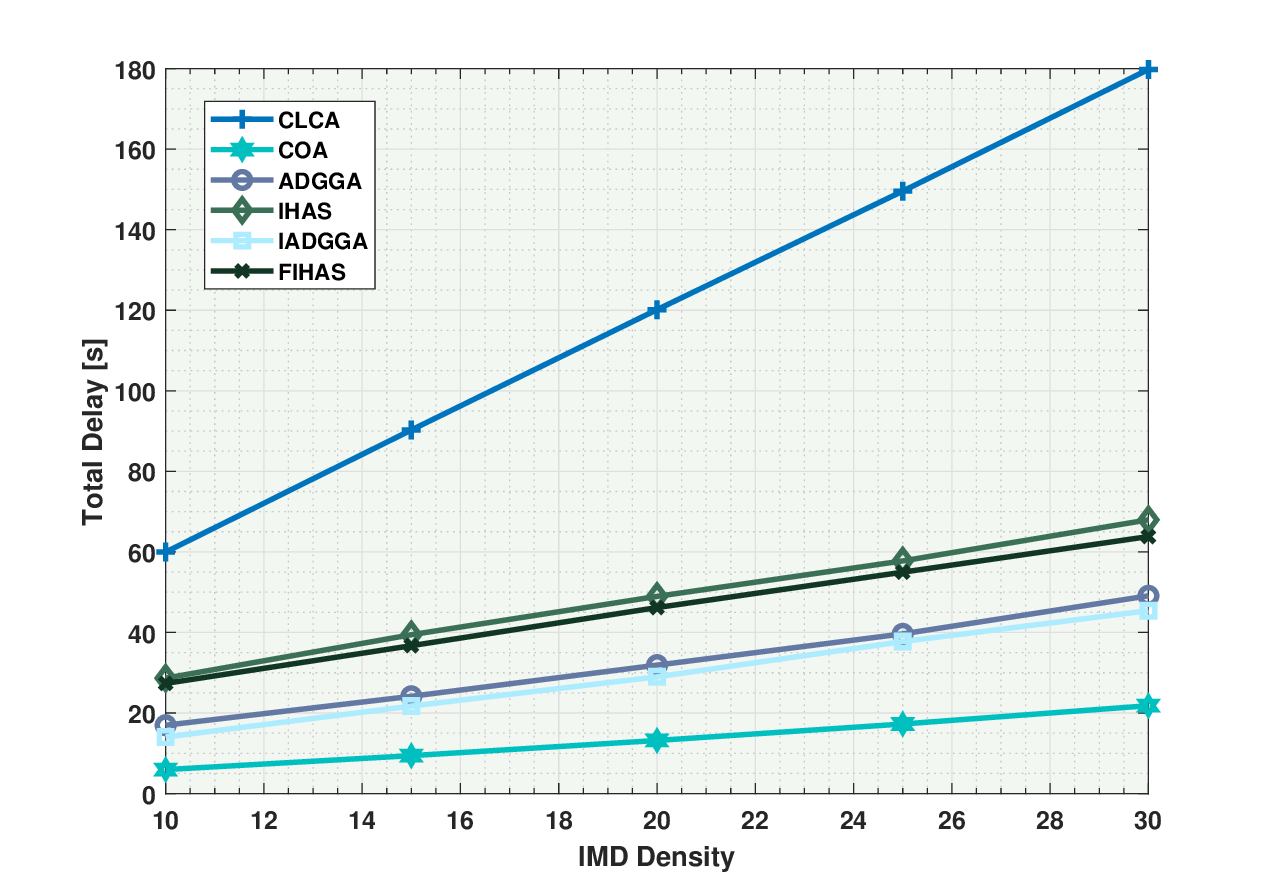}}
	\caption{Influences of IMD density on the total delay (TD).}
	\label{fig5}
\end{figure}
\par
Fig.\ref{fig5} shows the influences of IMD density on the total delay (TD), where TD refers to the sum of the time used for processing all tasks of all IMDs. As illustrated in Fig.\ref{fig5}, the TD should increase with IMD density since more IMDs often require more time to tackle their tasks. Among all offloading algorithms, COA may achieve the lowest TD since the best-channel association usually results in very low transmission time, but CLCA may achieve the highest TD since it always lets IMDs complete their tasks within maximum tolerable time. In addition, as revealed in Fig.\ref{fig5}, IADGGA may achieve a lower TD than ADGGA, and FIHAS may attain a lower TD than IHAS.
\begin{figure}[!t]
	\centerline{\includegraphics[width=3.8in]{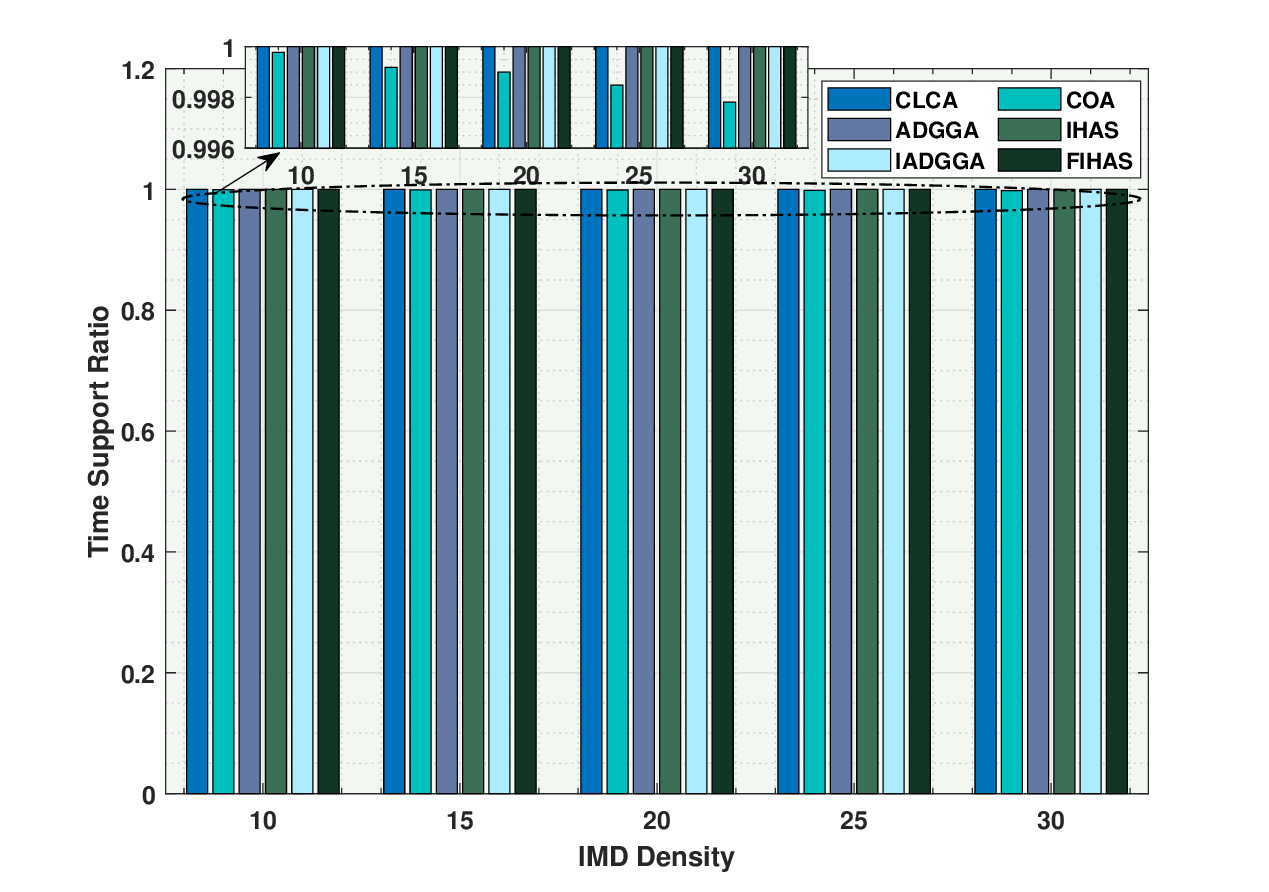}}
	\caption{Influences of IMD density on the time support ratio (TSR).}
	\label{fig6}
\end{figure}
\par
Fig.\ref{fig6} shows the influences of IMD density on the time support ratio (TSR), where TSR refers to the ratio of IMDs meeting the latency constraints to all IMDs. As illustrated in Fig.\ref{fig6}, under large enough penalty factors, the latency constraints of all IMDs can be strictly guaranteed in ADGGA, IADGGA, IHAS, and FIHAS. In addition, the latency constraints of all IMDs may also be strictly guaranteed in CLCA since it executes tasks within maximum tolerant time. However, the latency constraints of all IMDs cannot be strictly guaranteed in COA due to the extremely unbalanced load distribution and low resource utilization. Moreover, its TSR may generally decrease with the IMD density since more IMDs may result in more unbalanced load distribution and lower resource utilization.
\begin{figure}[!t]
	\centerline{\includegraphics[width=3.8in]{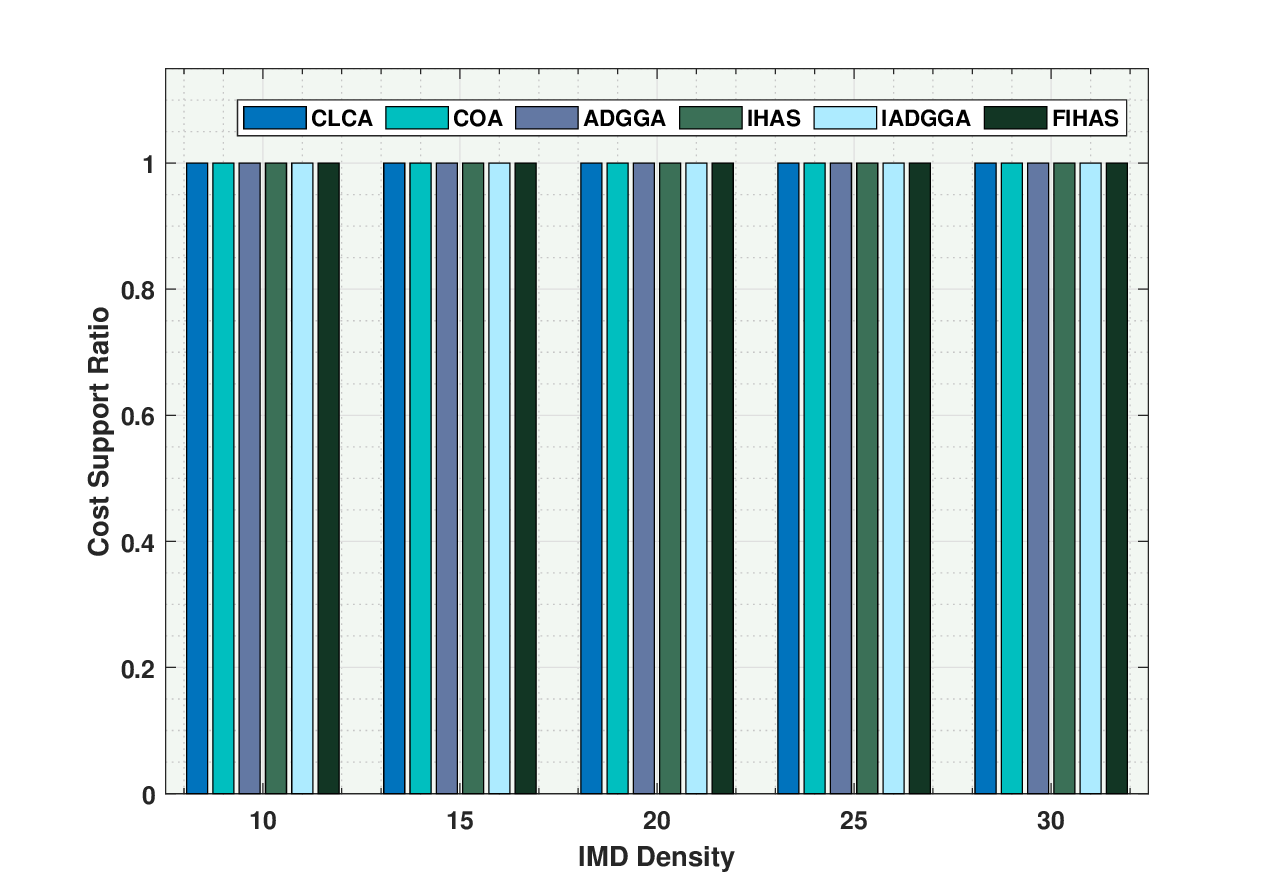}}
	\caption{Influences of IMD density on the cost support ratio (CSR).}
	\label{fig7}
\end{figure}
\par
Fig.\ref{fig7} shows the influences of IMD density on the cost support ratio (CSR), where CSR refers to the ratio of IMDs meeting the cost constraints to all IMDs. As illustrated in Fig.\ref{fig7}, under large enough penalty factors, the cost constraints of all IMDs can be strictly guaranteed in ADGGA, IADGGA, IHAS and FIHAS. In addition, the cost constraints of all IMDs may also be strictly guaranteed in CLCA since it has no tasks tackled at BSs. The cost constraints of all IMDs can also be strictly guaranteed in COA since it always selects cryptographic algorithms with minimal cost for each task.

\begin{figure}[!t]
	\centerline{\includegraphics[width=3.8in]{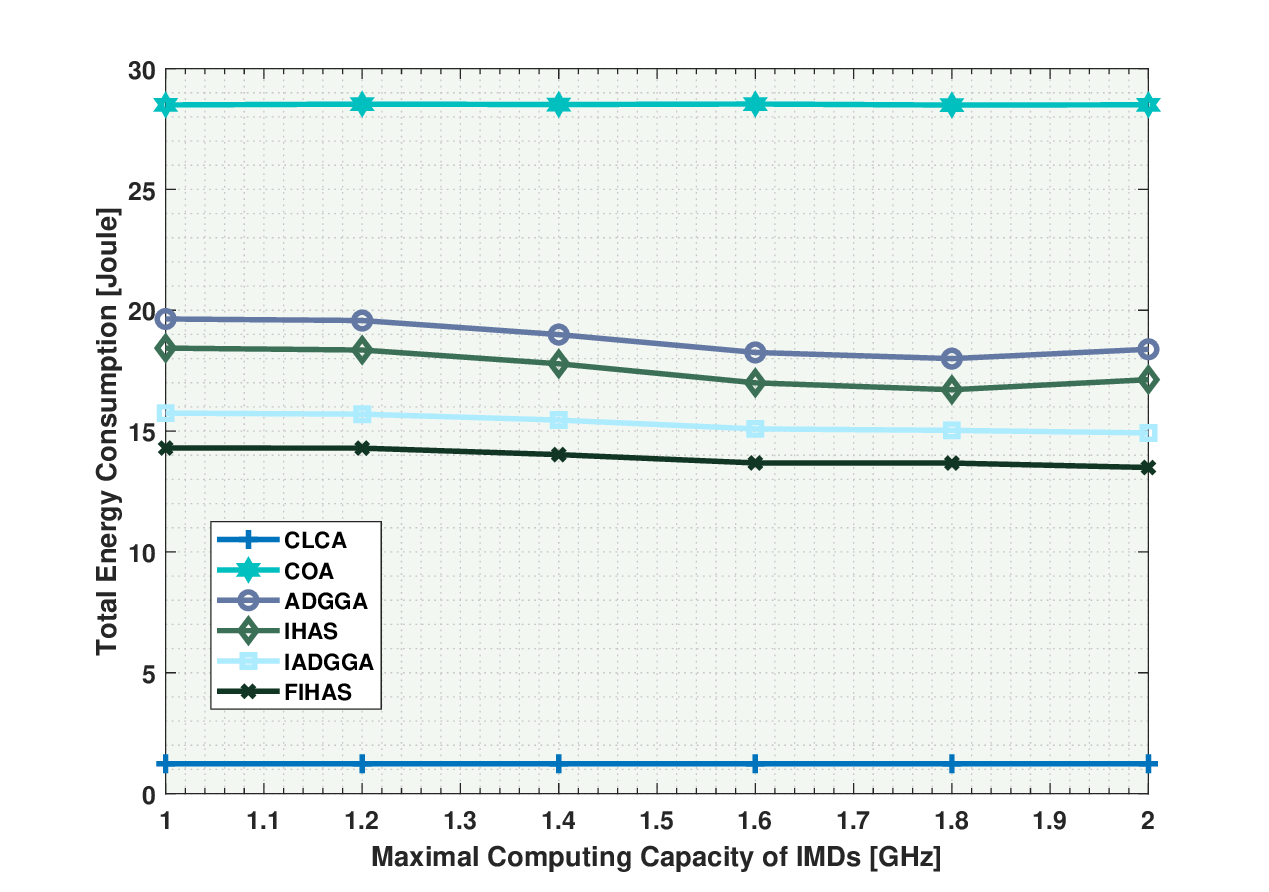}}
	\caption{Influences of maximal computing capacity of IMDs on TEC.}
	\label{fig8}
\end{figure}
\par
Fig.\ref{fig8} shows the influences of maximal computing capacity of IMDs (${f}_{k}^{\text{lmax}}={f}^{\text{max}}$ for any IMD $k$) on TEC. As illustrated in Fig.\ref{fig8}, besides CLCA and COA, the TEC of other algorithms may generally decrease with ${f}^{\text{max}}$ since the increased ${f}^{\text{max}}$ may result in the decreased local executing and encrypting time. In CLCA and COA, the TEC should not change with ${f}^{\text{max}}$. In CLCA, the tasks of IMDs are completed within the deadline, using computing capacity that is often less than maximal computing capacity; in COA, the local energy consumption has no relation to ${f}^{\text{max}}$. In addition, Similar to Fig.\ref{fig8}, COA may achieve the highest TEC, but CLCA may achieve the lowest TEC. In addition, IADGGA may achieve a lower TEC than ADGGA, and FIHAS may achieve a lower TEC than IHAS.
\begin{figure}[!t]
	\centerline{\includegraphics[width=3.8in]{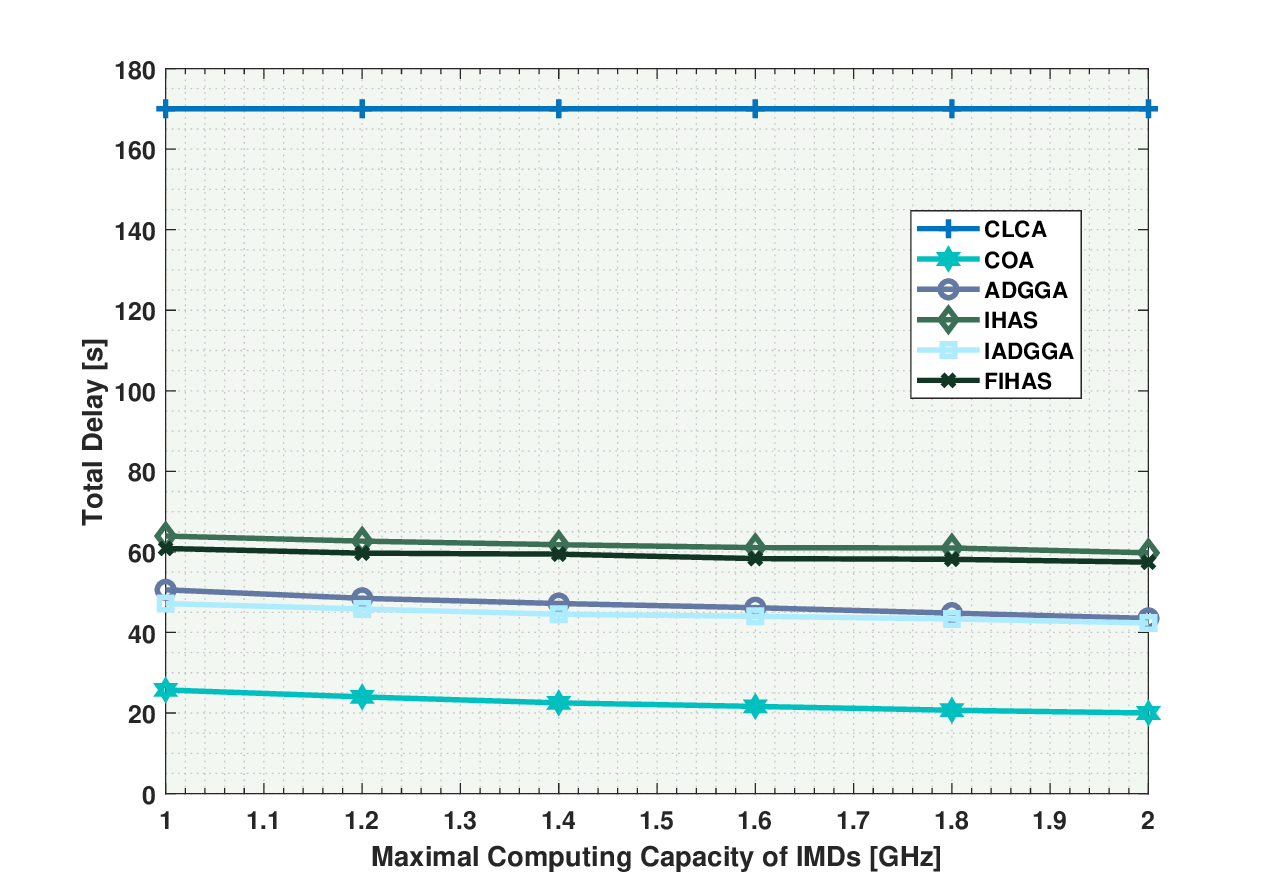}}
	\caption{Influences of maximal computing capacity of IMDs on TD.}
	\label{fig9}
\end{figure}
\par
Fig.\ref{fig9} shows the influences of ${f}^{\text{max}}$ (maximal computing capacity of IMDs) on TD. As illustrated in Fig.\ref{fig9}, besides CLCA, the TD of other algorithms may generally decrease with ${f}^{\text{max}}$ since the increased ${f}^{\text{max}}$ may result in the decreased local executing and encrypting time. In CLCA, the TEC should not change with ${f}^{\text{max}}$ since it has no locally executed tasks. Similar to Fig.\ref{fig5}, COA may achieve the lowest TD, but CLCA may achieve the highest TD. In addition, IADGGA may achieve a lower TD than ADGGA, and FIHAS may attain a lower TD than IHAS.
\begin{figure}[!t]
	\centerline{\includegraphics[width=3.8in]{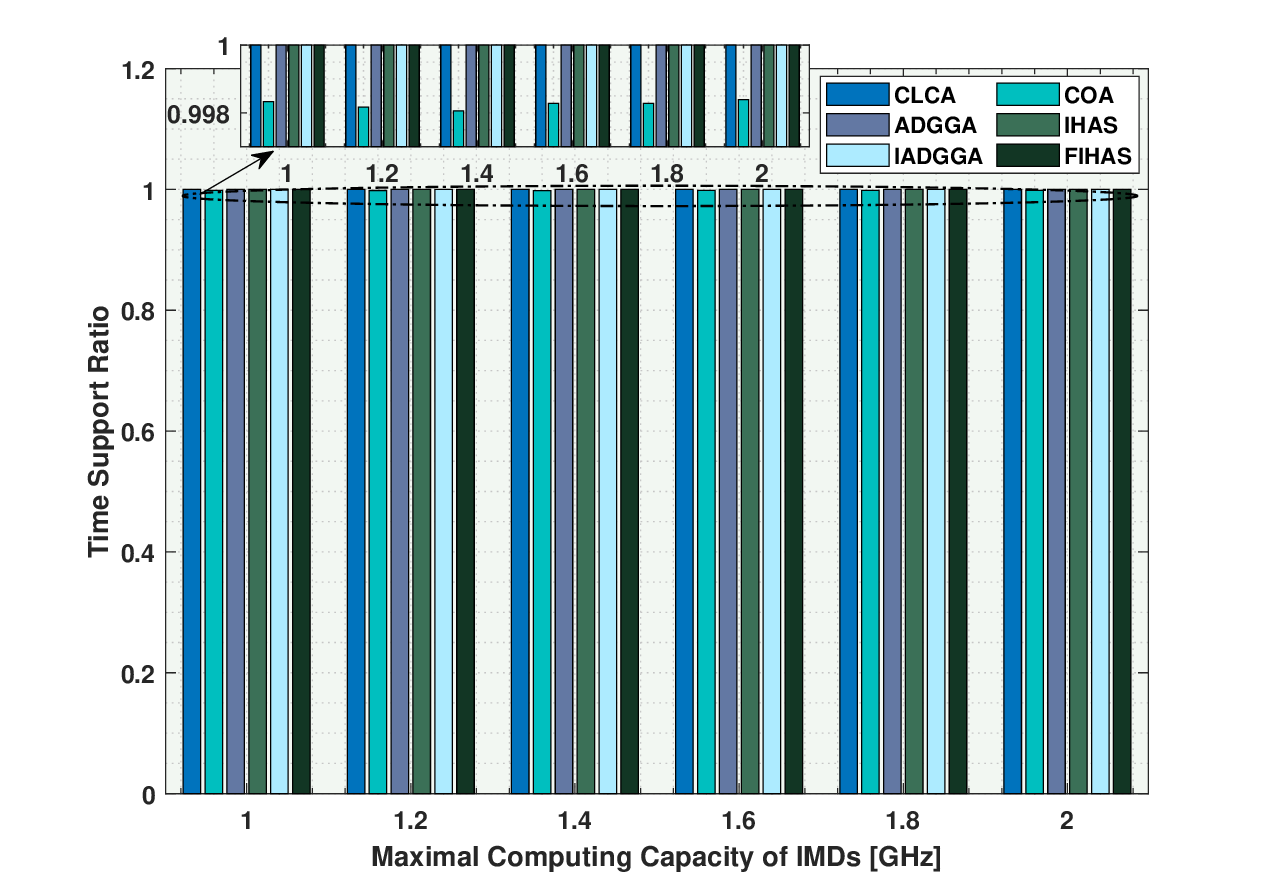}}
	\caption{Influences of maximal computing capacity of IMDs on TSR.}
	\label{fig10}
\end{figure}
\par
Fig.\ref{fig10} shows the influences of ${f}^{\text{max}}$ (maximal computing capacity of IMDs) on TSR. As revealed in Fig.\ref{fig6}, in Fig.\ref{fig10}, the latency constraints of all IMDs can be strictly guaranteed in ADGGA, IADGGA, IHAS, FIHAS, and CLCA. However, the latency constraints of all IMDs cannot be strictly guaranteed in COA. Moreover, its TSR may generally increase with ${f}^{\text{max}}$ since the increased ${f}^{\text{max}}$ results in decreased local executing and encrypting time.
\begin{figure}[!t]
	\centerline{\includegraphics[width=3.8in]{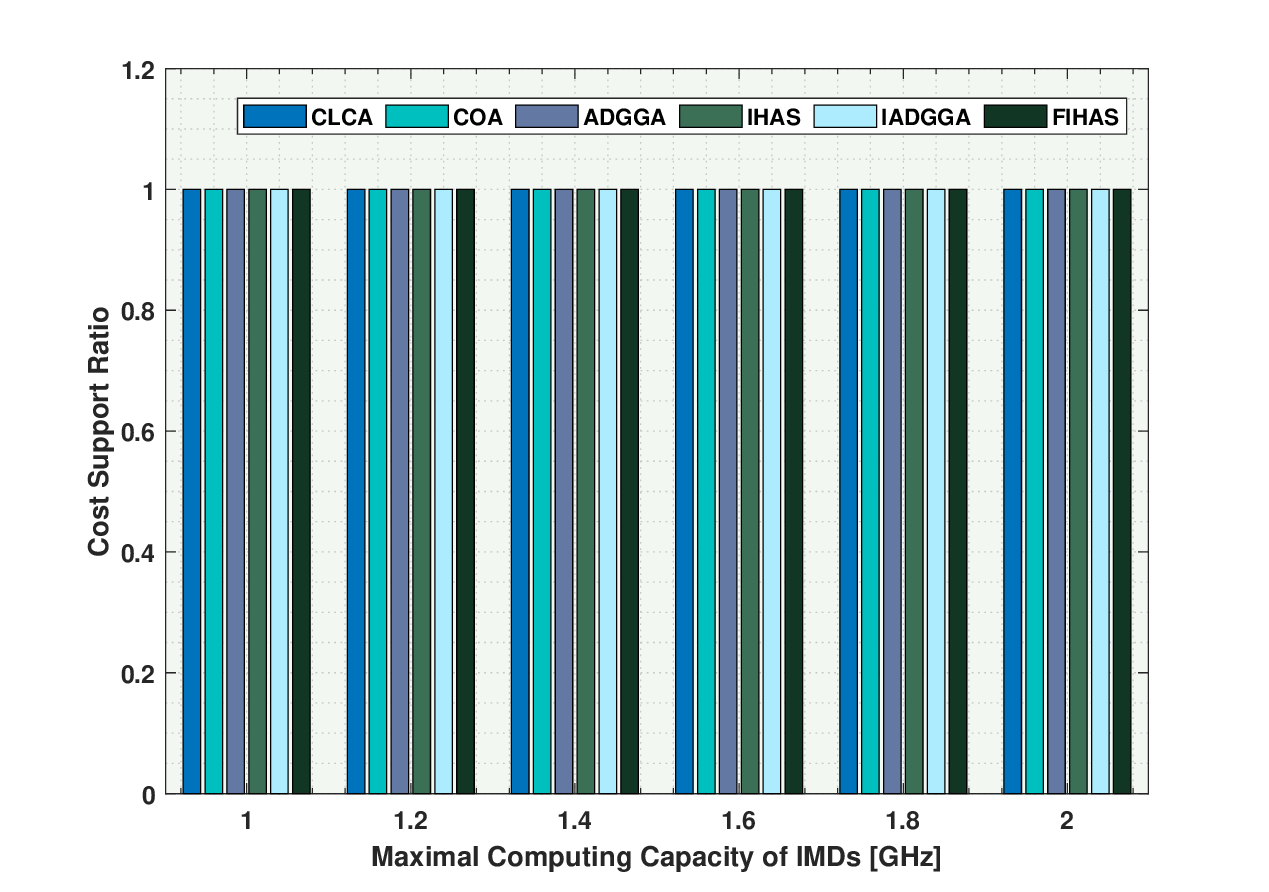}}
	\caption{Influences of maximal computing capacity of IMDs on CSR.}
	\label{fig11}
\end{figure}
\par
Fig.\ref{fig11} shows the influences of ${f}^{\text{max}}$ (maximal computing capacity of IMDs) on CSR. As revealed in Fig.\ref{fig7}, in Fig.\ref{fig11}, the cost constraints of all IMDs can be strictly guaranteed in ADGGA, IADGGA, IHAS, FIHAS, CLCA, and COA.
\begin{figure}[!t]
	\centerline{\includegraphics[width=3.8in]{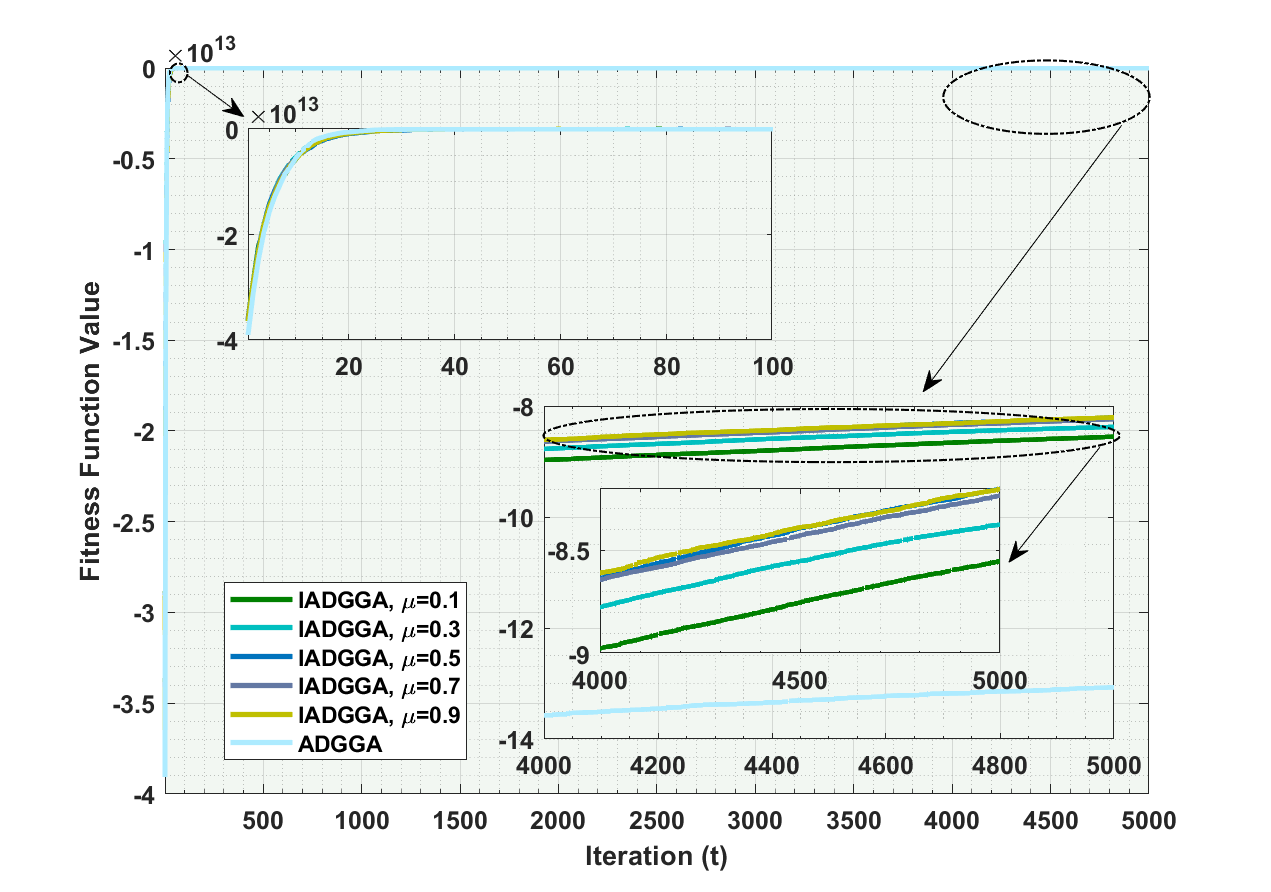}}
	\caption{Influences of similarity threshold on the fitness function value (FFV) and convergence.}
	\label{fig12}
\end{figure}
\par
Fig.\ref{fig12} shows the influences of similarity threshold ${\mu}_{3}$ on the fitness function value (FFV) and convergence. As illustrated in Fig.\ref{fig12}, both IADGGA and ADGGA can converge. ADGGA may converge faster than IADGGA since the latter adopts a two-step mutation to avoid premature convergence. Consequently, ADGGA may achieve a smaller FFV than IADGGA. Seen from Fig.\ref{fig12}, the FFV achieved by IADGGA may generally increase with ${\mu}_{3}$ since the individuals with higher similarity are eliminated and regenerated randomly, which can better guarantee the population diversity.
\begin{figure}[!t]
	\centerline{\includegraphics[width=3.8in]{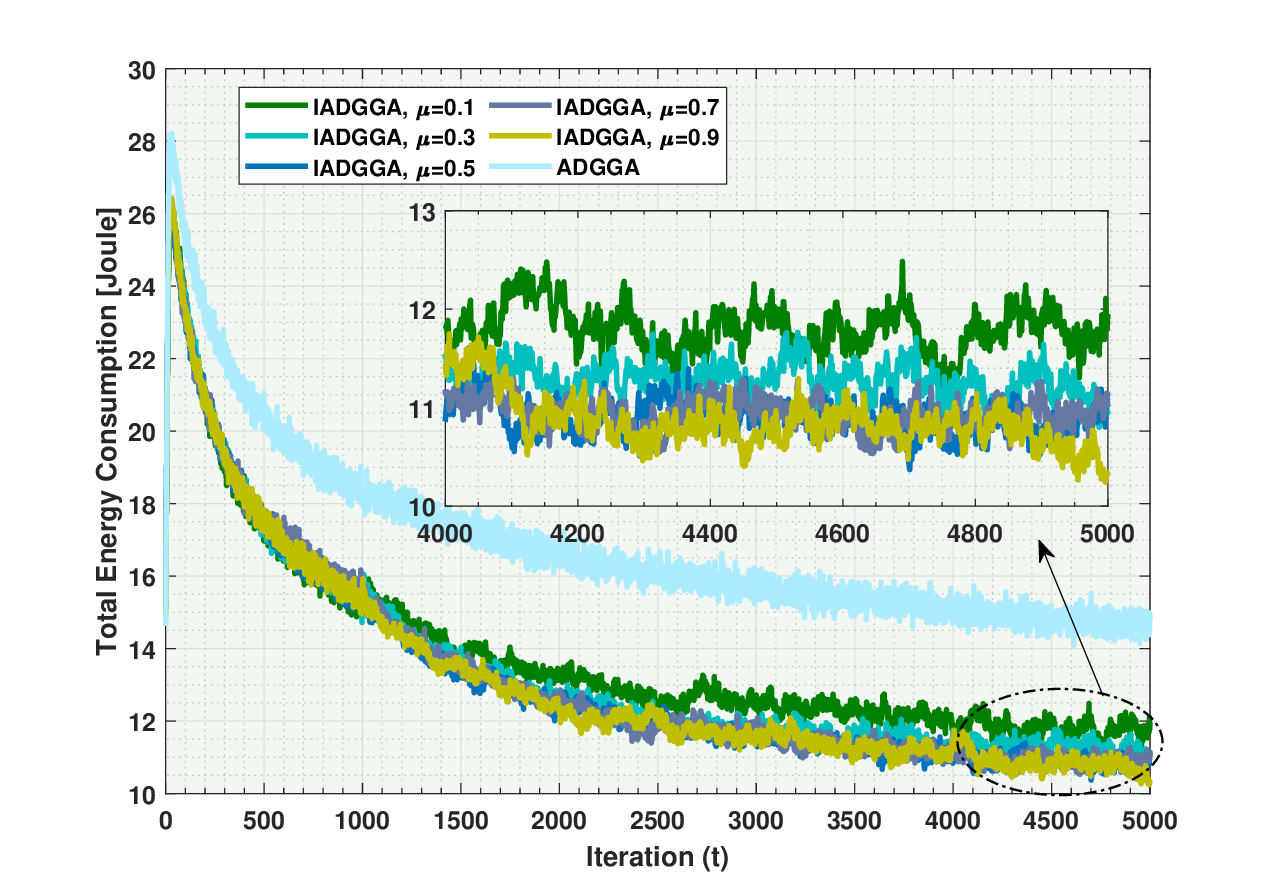}}
	\caption{Influences of similarity threshold on TEC and convergence.}
	\label{fig13}
\end{figure}
\par
Fig.\ref{fig13} shows the influences of similarity threshold ${\mu}_{3}$ on TEC and convergence. As revealed in Fig.\ref{fig12}, ADGGA may converge faster than IADGGA in Fig.\ref{fig13} since the latter adopts a two-step mutation to avoid premature convergence. Consequently, IADGGA may achieve lower TEC than ADGGA. As seen from Fig.\ref{fig13}, the TEC achieved by IADGGA may generally decrease with ${\mu}_{3}$ since some individuals with higher similarity and lower FFV are eliminated and regenerated randomly can better guarantee the population diversity.
\begin{figure}[!t]
	\centerline{\includegraphics[width=3.8in]{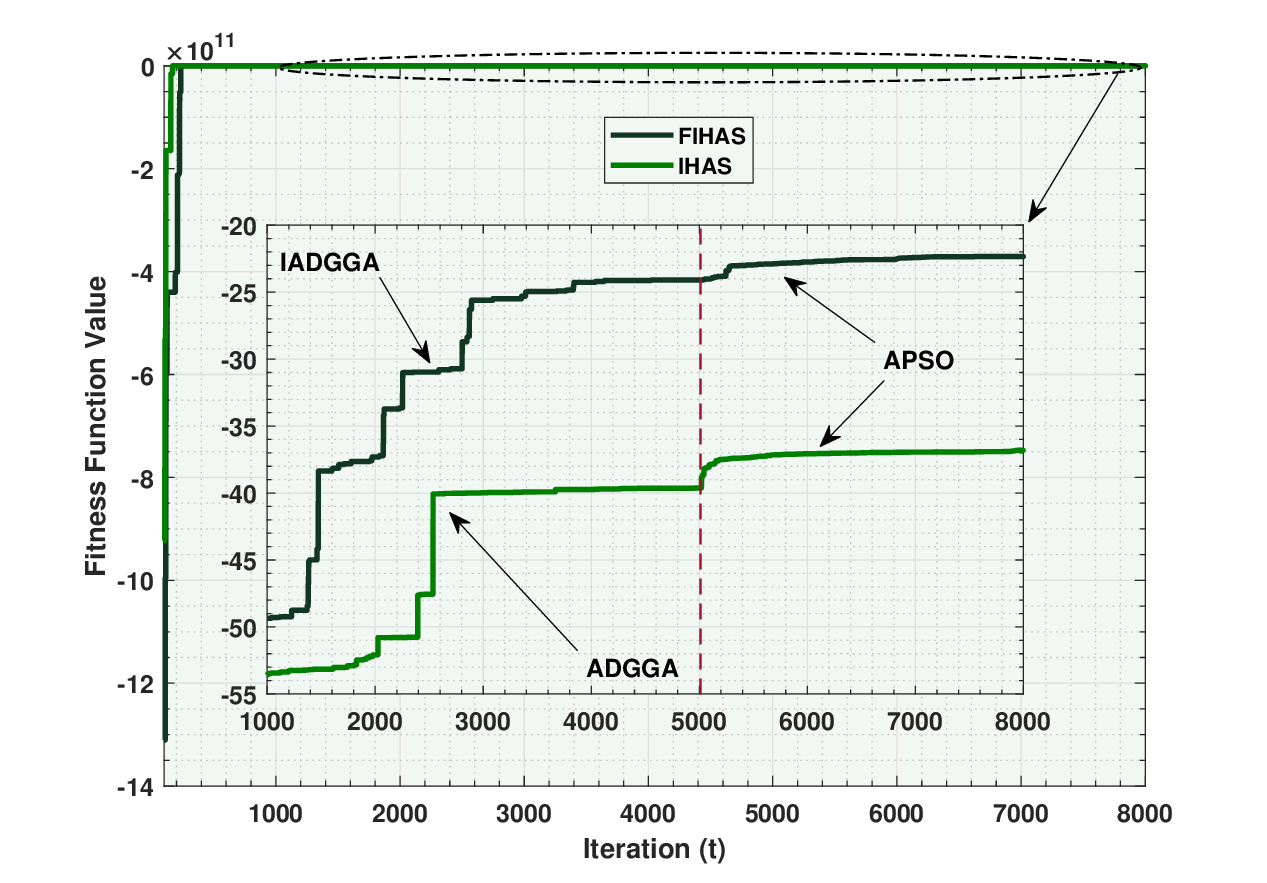}}
	\caption{Convergence comparison of IHAS and FIHAS.}
	\label{fig14}
\end{figure}
\par
Fig.\ref{fig14} shows the convergence of IHAS and FIHAS. As revealed in Fig.\ref{fig14}, FIHAS may achieve a higher FFV than IHAS since IADGGA (the first 5000 iterations) in the former has a stronger global search capability than ADGGA (the first 5000 iterations) in the latter. That is to say, ADGGA may have premature convergence. As shown in Fig.\ref{fig14}, APSO (the last 3000 iterations) in IHAS and FIHAS may further refine the solution achieved by ADGGA and IADGGA.
\section{Conclusion}\label{sec7}
In cache-assisted ultra-dense MEC networks featuring multi-slope channels, a combined approach involving NOMA, OFDMA, and BS clustering was adopted to alleviate network interferences and enhance frequency-spectrum efficiency. Given constraints related to task processing time, uplink power, local computing capacity, caching capacity, and security costs, we concurrently optimized device associations, power controls, caching, computation offloading, channel selections, security service assignments, and local computing capacity adjustments, aimed at minimizing the energy consumption of IMDs. To solve the formulated problem, IADGGA and FIHAS were devised. Simulation results indicate that, under stringent latency and cost constraints, IADGGA and FIHAS generally outperform other algorithms in terms of total energy consumption and delay. Potential avenues for future research include the development of additional swarm intelligence algorithms and the integration of blockchain technology.
\vspace{-0.5in}
\begin{IEEEbiography}[{\includegraphics[width=1in,height=1.1in,clip,keepaspectratio]{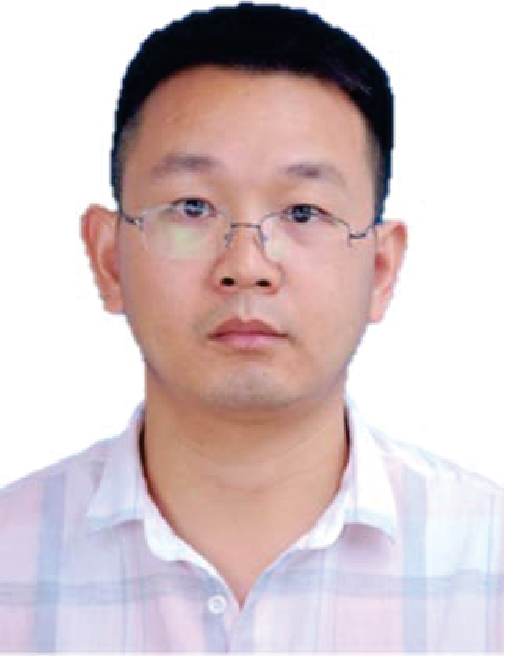}}]{Tianqing Zhou}
 received the Ph.D. degree in Information and Communication Engineering from Southeast University, Nanjing, China, in 2016. He is currently an associate Professor with the School of Information and Software Engineering, East China Jiaotong University. He is the author or coauthor of more than 40 journal papers indexed by SCI, and holds more than 10 patents. His current research interests include mobile edge computing and caching, artificial intelligence algorithms, wireless resource management and ultra-dense networks.
\end{IEEEbiography}
\vspace{-0.5in}
\begin{IEEEbiography}[{\includegraphics[width=1in,height=1.1in,clip,keepaspectratio]{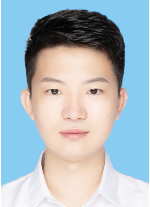}}]{Bobo Wang}
 received the M.E. degree in Information and Communication Engineering at East China Jiaotong University, Nanchang, China, in 2024. He is currently a lecturer with the College of Electronic and Information Engineering, Huaibei Institute of Technology. His current research interests include mobile edge computing and ultra-dense networks.
\end{IEEEbiography}
\vspace{-0.65in}
\begin{IEEEbiography}[{\includegraphics[width=1in,height=1.2in,clip,keepaspectratio]{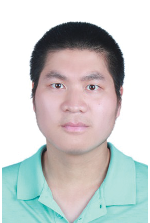}}]{Dong Qin} received the Ph.D. degree in Information and Communication Engineering from Southeast University, Nanjing, China, in 2016. He is currently an associate Professor with the School of Information Engineering, Nanchang University. His current research interests lie in the area of cooperative communication and OFDM techniques.
\end{IEEEbiography}
\vspace{-0.6in}
\begin{IEEEbiography}[{\includegraphics[width=1in,height=1.1in,clip,keepaspectratio]{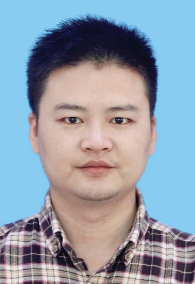}}]{Xuefang Nie}
 received the Ph.D. degree in Information and Communication Engineering from the Harbin Institute of Technology, China, in 2018. He is currently an associate Professor with the School of Information and Software Engineering, East China Jiaotong University. His research interests include heterogeneous network, wireless communications, stochastic geometry, mobile edge computing, fog computing and IoT technology.
\end{IEEEbiography}
\vspace{-0.6in}
\begin{IEEEbiography}[{\includegraphics[width=1in,height=1.1in,clip,keepaspectratio]{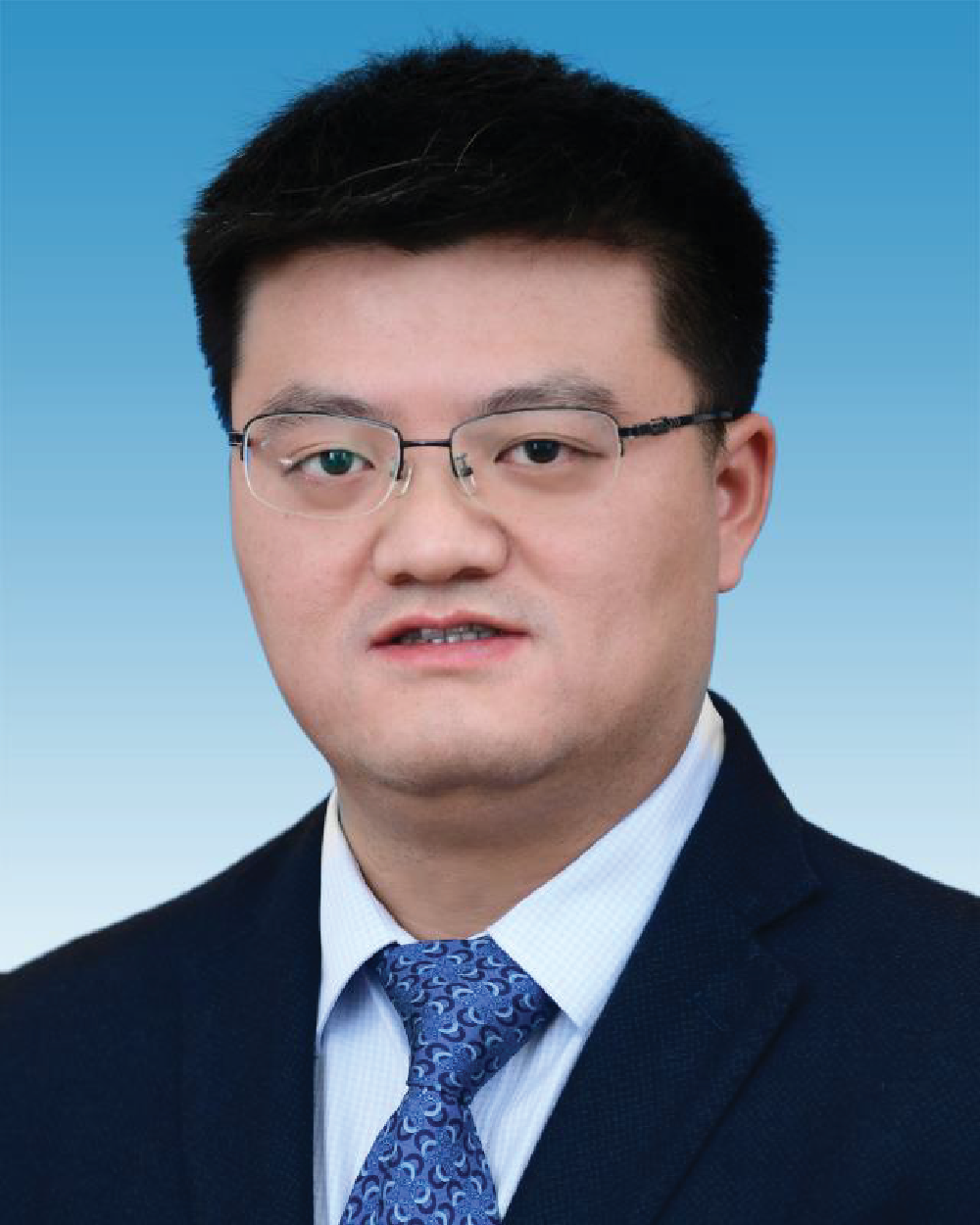}}]{Nan Jiang}
received the PhD degree in Nanjing University of Aeronautics and Astronautics, Nanjing, China, in 2008. Currently he is a professor in School of Information and Software Engineering, East China Jiaotong University. He was the research scholar in Complex Networks and Security Research Lab at Virginia Tech between 2013 and 2014. His current research interests lie in wireless sensor networks, the scalable sensor networks, wireless protocol and architecture, distributed computing.
\end{IEEEbiography}
\vspace{-0.6in}
\begin{IEEEbiography}[{\includegraphics[width=1in,height=1.2in,clip,keepaspectratio]{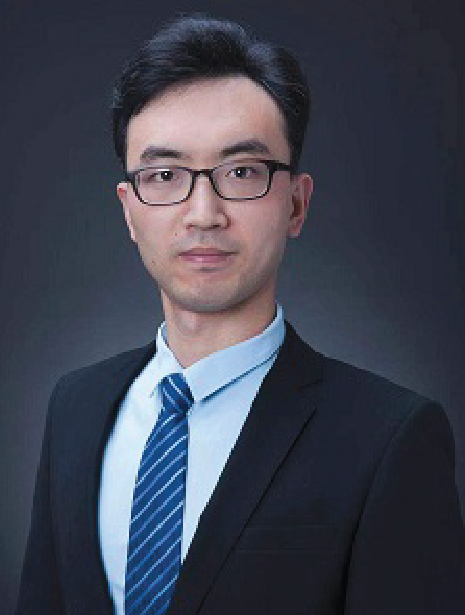}}]{Chunguo Li}
 (SM'16) received the bachelor's degree in Wireless Communications from Shandong University in 2005, and the Ph.D. degree in Wireless Communications from Southeast University in 2010. In July 2010, he joined the School of Information Science and Engineering, Southeast University, Nanjing China, where he is currently an Advisor of Ph.D candidates and Full Professor. From June 2012 to June 2013, he was the Post-Doctoral researcher with Concordia University, Montreal, Canada. From July 2013 to August 2014, he was with the DSL laboratory of Stanford University as visiting associate professor. From August 2017 to July 2019, he was the adjunct professor of Xizang Minzu University under the supporting Tibet program organized by China National Human Resources Ministry. He is the Fellow of IET, the Fellow of China Institute of Communications (CIC), and IEEE CIS Nanjing Chapter Chair. His research interests are in cell-free distributed MIMO wireless communications for 6G, and machine learning based image/video signal processing algorithm design.
\end{IEEEbiography}





\end{document}